\documentclass[11pt,a4paper]{article}
\usepackage{jheppub}

\usepackage{enumerate}
\usepackage{amsmath}
\usepackage{amsfonts}
\usepackage{amssymb}
\usepackage[utf8]{inputenc}
\usepackage[T1]{fontenc}
\usepackage{mathtools}
\usepackage{wasysym}
\usepackage{accents}
\usepackage[normalem]{ulem}
\usepackage[svgnames]{xcolor}
\usepackage[colorlinks=true,linkcolor=blue]{hyperref}
\DeclareMathOperator{\sech}{sech}

\DeclareMathOperator{\arcsinh}{arcsinh}

\begin{document}
\newcommand{\Areia}{
\affiliation[a]{Department of Chemistry and Physics, Federal University of Para\'iba, Rodovia BR 079 - Km 12, 58397-000 Areia-PB,  Brazil.}
}
\newcommand{\Lavras}{
\affiliation[b]{Physics Department, Federal University of Lavras, Caixa Postal 3037, 37200-000 Lavras-MG, Brazil.}
}

\newcommand{\JP}{
\affiliation[d]{Physics Department, Federal University of Para\'iba, \\Caixa Postal 5008, 58059-900, Jo\~ao Pessoa, PB, Brazil.}
}

% \title{Particle decay phenomenology from Planck scale deformed relativity}
%\title{Corrected particle distributions from Planck scale deformed relativity}
%\title{Planck scale relativistic amplified corrections to time dilation and particle distributions}
%\title{Planck scale relativistic amplified corrections to particle decay phenomenology}
%\title{Corrections for time dilation and particle distribution from deformed relativity at the Planck scale}
%\title{Particle distributions from Planck scale deformed relativity}

\title{Two-body decays in deformed relativity}
%new
%\title{Phenomenological opportunities of two-body decays in deformed relativity}
%\title{Two-body decays in deformed relativity}
%\title{Opportunities of Finsler-inspired deformed relativity in two-body decays}
%title{Phenomenology of two-body decays in Finsler-inspired deformed relativity}

\author[a,b]{Iarley P. Lobo,}
\emailAdd{lobofisica@gmail.com}

\author[c]{Christian Pfeifer,}
\emailAdd{christian.pfeifer@ut.ee}

\author[d]{Pedro H. Morais,}
\emailAdd{phm@academico.ufpb.br}

\author[e]{Rafael {Alves Batista},}
\emailAdd{rafael.alvesbatista@uam.es}

\author[d]{Valdir B. Bezerra}
\emailAdd{valdir@fisica.ufpb.br}

\affiliation[a]{Department of Chemistry and Physics, Federal University of Para\'iba, Rodovia BR 079 - Km 12, 58397-000 Areia-PB,  Brazil.}
\affiliation[b]{Physics Department, Federal University of Lavras, Caixa Postal 3037, 37200-000 Lavras-MG, Brazil.}
\affiliation[c]{ZARM,Am Fallturm 2, University of Bremen, 28359 Bremen, Germany.}
\affiliation[d]{Physics Department, Federal University of Para\'iba, Caixa Postal 5008, 58059-900, Jo\~ao Pessoa, PB, Brazil.}
\affiliation[e]{Instituto de F\'isica Te\'orica UAM-CSIC, C/ Nicol\'as Cabrera 13-15, 28049 Madrid, Spain.}
\date{\today}

\abstract{
Deformed relativistic kinematics is a framework which captures effects, that are expected from particles and fields propagating on a quantum spacetime, effectively. They are formulated in terms of a modified dispersion relation and a modified momentum conservation equation. In this work we use Finsler geometry to formulate deformed relativistic kinematics in terms of particle velocities. The relation between the Finsler geometric velocity dependent formulation and the original momentum dependent formulation allows us to construct deformed Lorentz transformations between arbitrary frames. Moreover, we find the corresponding compatible momentum conservation equation to first order in the Planck scale deformation of special relativity based on the $\kappa$-Poincar\'e algebra in the bicrossproduct basis. We find that the deformed Lorentz transformations, as well as the deformed time dilation factor, contain terms that scale with the energy of the particle under consideration to the fourth power. We derive how the distributions of decay products are affected when the deformed relativity principle is satisfied and find, for the case of a pion decaying into a neutrino and a muon, that the ratio of expected neutrinos to muons with a certain energy is just slightly modified when compared to the predictions based on special relativity. We also discuss the phenomenological consequences of this framework for cosmic-ray showers in the atmosphere.}

\arxivnumber{2112.12172}

\maketitle

%%%%%%%%%%%%%%%%%%%%%%%%%%%%%%%%%%%%%%%%%%%%%%%%%%%%%%%%%%%%%%%%%%
\section{Introduction}\label{sec:intro}

The search for the nature of spacetime when gravitational degrees of freedom are quantized has proven to be one of the most challenging quests of physical sciences. Many theoretical attempts have been pursued in the last decades \cite{Rovelli:1997yv,Mukhi:2011zz,Loll:1998aj}, which allowed the emergence of phenomenological proposals that gather some in common and observationally compelling properties of these approaches in a unified way that can be confronted with present or near future observations \cite{AmelinoCamelia:2008qg}, specially in the multi-messenger astronomy era \cite{Addazi:2021xuf}.
\par
Among these proposals, a prominent role is played by the those that predict a {\it violation} or {\it deformation} of the special relativistic principles (referred as LIV and DSR scenarios, respectively) in the regime in which $\hbar\rightarrow 0$ and $G\rightarrow 0$, but that preserves the Planck energy scale $E_{\text{P}}=\sqrt{c^5\hbar/G}\approx 1.22\times 10^{19}\, \text{GeV}$. The phenomenological opportunities of this approach range among several areas, for instance threshold effects in particle interactions \cite{Jacobson:2002hd,Carmona:2020whi,Li:2021tcw}, time delays in the arrivals of relativistic particles with different energies due to modified dispersion relations (MDRs) \cite{Levy:2021oec,Ellis:2002in,Jacob:2008bw,Rosati:2015pga,Barcaroli:2016yrl,Pfeifer:2018pty,Terzic:2021rlx}, gravitational lensing observations \cite{Glicenstein:2019rzj,Barcaroli:2017gvg}, violation of CPT symmetry \cite{Klinkhamer:2003ec,Arzano:2019toz}, etc (we refer the reader to Refs.~\cite{AmelinoCamelia:2008qg,Liberati:2013xla,Addazi:2021xuf} for reviews in modern searches of these effects).
\par
High-energy astrophysical messengers such as cosmic rays, gamma rays, and neutrinos have been widely used to search for signatures of quantum gravity (QG). Most investigations focused on searches for LIV signatures, both in the propagation of these cosmic messengers and in the atmospheric showers they induce (see, e.g.,~\cite{Martinez-Huerta:2020cut, AlvesBatista:2019tlv, Arguelles:2021kjg, Addazi:2021xuf, Wei:2021vvn} for reviews). Nevertheless, their potential for probing DSR theories have not yet been fully exploited, despite this being a clearly promising avenue for investigating QG phenomenology. One remarkable example of a problem whose solution might potentially hint at QG is the so-called muon puzzle, which consists in an $8\sigma$ excess in the measured number of muons in air showers initiated by ultra-high-energy cosmic rays (UHECRs) compared with theoretical predictions~\cite{PierreAuger:2014ucz, PierreAuger:2016nfk, PierreAuger:2021qsd, Soldin:2021wyv, Dembinski:2021szp}. In this case, the muon puzzle could be a consequence of the changes in the kinematics associated with the decay of the particles that compose the shower.
\par
Recently, in approaches that deform relativistic symmetries, the lifetime of fundamental particles was shown to be an experimentally compelling observable with the potential of reaching the Planck scale in the near future \cite{Arzano:2019toz,Arzano:2020rzu,Lobo:2020qoa}. In \cite{Lobo:2020qoa} we used Finsler geometry (see for example \cite{Pfeifer:2019wus} for an overview on Finsler geometry) to construct the geometric clock (the length measure for worldliness) which is compatible with DSR symmetries, and derived the dilated lifetime of particles. The corrections to the special relativistic time dilation are interpreted as arising from the effective spacetime probed by particles which are subject to kinematic corrections due to propagation on a quantum spacetime.

One of the key results that emerged in \cite{Lobo:2020qoa} consists in the identification of a deformed Lorentz transformation that relates the (4-)momenta of a particle in its rest frame to the lab frame simply from the relation between the momenta and the (4-)velocity derived by the Finsler function. Most interestingly, the first order Planck-scale correction is governed by the fourth power of the Lorentz factor. This dominant contribution is of high importance since it works as an amplifier to the Planck scale that can turn the lifetime of fundamental particles in accelerators as an observable that could be measured with Planck scale sensitivity.
\par
We wish to diversify the analysis of observables which are derived from the transformations between different frames which move with a certain velocity relatively to each other, and/or seek a more complete description that is compatible with the relativistic principle. Therefore, it becomes necessary to find the deformed Lorentz transformations that connect arbitrary frames.  In addition, since we work in a DSR approach, it becomes fundamental to investigate the composition law of momenta in a Finsler-DSR-compatible way \cite{Amelino-Camelia:2000stu}, in order to preserve the nature of relativistic interactions for every inertial frame.
\par
Previous investigations have focused on an infinitesimal version of the symmetry transformations, i.e., without integrating the symmetry transformation parameters \cite{Amelino-Camelia:2014rga}. Here we construct the finite version of the deformed Lorentz transformation and its related composition law, in the sense that the transformation parameter which connects different frames does not need to be small; this is what we call a {\it{finite transformation}}. However, this should not be confused with the fact that we investigate first order in Planck-scale corrections, i.e.\ first order Finsler deviations from Minkowski spacetime which emerge from first order in Planck-scale deformed symmetries. The reason for this choice consists in the positive outcome due to the amplifying factor that grows with the fourth-power in the Lorentz factor that might bring us close to the Planck scale with accelerators or astrophysical observations. Investigating these issues is the main purpose of this paper.
\par
The paper is organized as follows: In Section~\ref{sec:fin_def}, we derive the finite deformed Lorentz transformation between arbitrary frames in the bicrossproduct basis $\kappa$-Poincar\'e-inspired Finsler geometry. In Section~\ref{sec:comp_law}, we derive the Finsler-DSR-compatible composition law of energy and momentum and discuss its ambiguities. In Section~\ref{sec:particle_decay}, we discuss the implications of these results for some equations that involve particle decays. In Section~\ref{sec:conc}, we draw our conclusions and prospects.

%%%%%%%%%%%%%%%%%%%%%%%%%%%%%%%%%%%%%%%%%%%%%%%%%%%%%%%%%%%%%%%%%%
\section{Deformed Lorentz transformations from Finsler geometry}\label{sec:fin_def}

Modified relativistic kinematics lead to a momentum dependent spacetime geometry which can be captured by a non-trivial geometry phase space (the cotangent bundle) of spacetime \cite{Barcaroli:2015xda,Carmona:2019fwf,Pfeifer:2021tas}. For applications, it is useful to reformulate the momentum-dependent spacetime geometry in a dual way in terms of a velocity-dependent geometry on the tangent bundle of spacetime, which leads to the mathematical framework of Finsler geometry \cite{Pfeifer:2019wus}.

Here, we assume as our working model the so called bicrossproduct basis-inspired Finsler geometry discussed in \cite{Amelino-Camelia:2014rga,Lobo:2020qoa}. In this case, the Finsler function in the Cartesian coordinate system,  calculated from a Helmholtz action with a MDR is given by
\begin{equation}\label{f_function1}
    F(\dot{x})=\sqrt{\eta(\dot{x},\dot{x})}-\frac{\ell m}{2}\frac{\dot{x}^0\delta_{ij}\dot{x}^i\dot{x}^j}{\eta(\dot{x},\dot{x})}\, ,
\end{equation}
where $\eta(\dot x,\dot x) = \eta_{\mu\nu}\dot x^\mu \dot x^\nu$ is the Minkowski norm of the velocity $4$-vector $\dot{x}$, $\delta_{ij}$ are the components of the $3$-dimensional euclidean metric, lower Latin indices span space coordinates, ``$\cdot$'' is the derivative with respect to the particle's worldline parameter, and $m$ is the mass of the particle that ``probes'' the Finsler spacetime. We define $\ell \doteq \kappa^{-1}$, where $\kappa$ is the energy scale of the deformation (the scale that defines the $\kappa$-Poincar\'e algebra), expected to be of the order of the Planck energy $E_{\text{P}}\approx 1.2\times 10^{19}\, \text{GeV}$. We assume natural units, i.e., $c=\hbar=1$. 

Physically, the Finsler function defines the geometric clock via the length measure for worldlines~$x(\tau)$
\begin{align}\label{clock}
    S[x] = \int\ d\tau\ F(\dot x)\,.
\end{align}
From the Finsler function \eqref{f_function1} it can be verified that the particle's momentum satisfy the following MDR that is functionally equivalent to the Cartesian coordinates realization of the Casimir operator of the $\kappa$-Poincar\'e algebra in the bicrossproduct basis \cite{Amelino-Camelia:2014rga}
\begin{equation}\label{MDR1}
    m^2=H(p) = g^{\text{F}\mu\nu}\left(\dot{x}(p)\right)p_{\mu}p_{\nu}=p_0^2-\delta^{ij}p_ip_j-\ell p_0\delta^{ij}p_ip_j\, .
\end{equation}
In the above equation, $g^{\text{F}\mu\nu}$ are the contravariant components of the Finsler metric, $g_{\mu\nu}^{\text{F}}=\partial^2 (F^2/2)/\partial \dot{x}^{\mu}\partial \dot{x}^{\nu}$, and the physical momentum $p_{\mu}$, which satisfy the dispersion relation, are defined directly from $F$ as follows:
\begin{equation}\label{eq:p_mu1}
    p_{\mu}(\dot x)=m\frac{\partial F}{\partial \dot{x}^{\mu}}\,.
\end{equation}
This last observation, which was demonstrated in \cite{Lobo:2020qoa}, is an important step towards implementing all DSR symmetries and structures on velocity space, which are usually constructed in terms of deformed algebras on momentum space. In particular \eqref{eq:p_mu1} implements deformed Lorentz transformations, as we will recall next. Still work in progress is the self consistent construction of a modified addition law of velocities which emerges from the modified addition law of momenta used in DSR theories.

\subsection{Lorentz transformations between rest frame and laboratory frame}
In calculating the momenta from \eqref{f_function1} and parametrizing the particle's worldline with the coordinate time $x^0$, we observe that the energy $E = p_0$ and spatial momentum $p_i$ are linked to the mass of the particle by the following relation:
\begin{align}
&p_0(v)=\gamma m-\frac{\ell}{2}m^2(\gamma^2-1)(2\gamma^2-1)\, ,\label{transf_0}\\
&p_i(v)=-v_i \gamma m+\ell m^2 v_i \gamma^4\, ,\label{transf_1}
\end{align}
where $v^i=dx^i/dx^0$, $v^2=\delta_{ij}v^{i}v^{j}$, and $\gamma^{-1}=\sqrt{1-v^2}$ is what we call the velocity Lorentz factor given in terms of the particle's velocity as measured by the laboratory (lab) frame. Notice that Eqs.~(\ref{transf_0}) and (\ref{transf_1}) are, in fact, displaying the 4-momentum $p_\mu$ (energy and spatial momentum) of the particle in the lab frame as a function of the particle's velocity in this frame (since we are using the coordinate time $x^0$ as curve parameter) and its mass, which is nothing but the particle's energy in its rest frame ($p_0(0)=m$). When $\ell\rightarrow 0$, we consistently recover the usual Lorentz transformations between the particle's rest frame and a lab frame, which are a consequence of the symmetries of the underlying Minkowski spacetime.

%Also notice that, when $\ell\rightarrow 0$, one is actually expressing the Lorentz transformation that relates the energy/momentum of a particle in its rest frame to the lab frame in which it moves with velocity $v$ and the rest frame, which is, in fact a symmetry of the Minkowski spacetime. 

As we verified in \cite{Lobo:2020qoa}, Eqs.~(\ref{transf_0}) and (\ref{transf_1}) are not only a transformation between quantities in the rest and the lab frames, but they also constitute a symmetry of the Finsler spacetime, in the sense that the dispersion relation $H(p) = H(\tilde{p}) = m^2$ is preserved. In other words, for a particle at rest ($v=0,\gamma=1$), the dispersion relation implies $p_0 = m$ and $p_i = 0$. If we consider $p_{\mu}$ as function of $\gamma$ and apply the transformation 
\begin{align}\label{eq:pofgamma}
    p_0=p_0(1) \to \tilde{p}_0 = p_0(\gamma) = m \frac{\partial}{\partial \dot x^0}F \textrm{ and } p_i=p_i(1)\to \tilde{p}_i= p_i(\gamma) = m \frac{\partial}{\partial \dot x^i}F\, ,
\end{align}
then $\tilde p_\mu$ also satisfies the dispersion relation. This implies that the transformations just highlighted are, in fact, canonical \textit{deformed Lorentz transformations} between the rest and the lab frames, which emerge from the underlying spacetime geometry.

\begin{figure}[h!]
    \centering
    \includegraphics[scale=0.7]{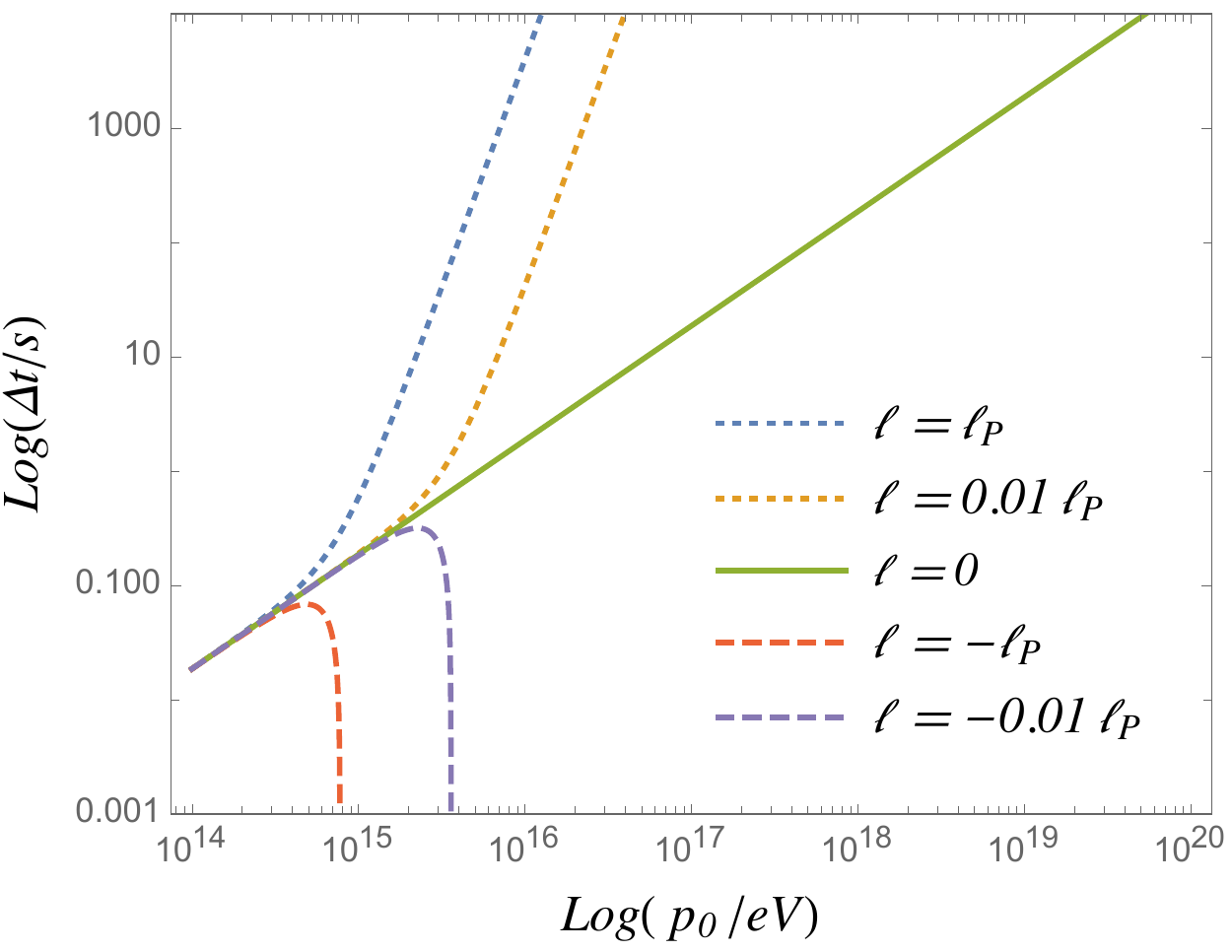}\caption{Lifetime of pion $\pi^{\pm}$ ($m=140\, \text{MeV}$), according to equation \eqref{eq:lifetime} following $\gamma_{\text{DSR}}$.}\label{lifetime1}
\end{figure}

Eq.~(\ref{transf_0}) also connects what we call the velocity Lorentz factor $\gamma$ and the momentum Lorentz factor $\bar{\gamma}\doteq p_0/m$ as $\gamma=\bar{\gamma}+\frac{\ell m}{2}(1-3\bar{\gamma}^2+2\bar{\gamma}^4)$, which, along with the formulation of the clock postulate in Finsler geometry \cite{Lobo:2020qoa}, gives the following prediction for the dilated lifetime of fundamental particles as measured in the lab frame
\begin{equation}\label{eq:lifetime}
    \Delta t=\frac{p_0}{m}\Delta \tau\left[1+\frac{\ell}{2}\left(\frac{m^2}{p_0}-2p_0+\frac{p_0^3}{m^2}\right)\right]\doteq\gamma_{\text{DSR}}\Delta \tau\, ,
\end{equation}
wherein $\Delta \tau$ corresponds to the lifetime of a fundamental particle in its rest frame. We defined a modified Lorentz factor $\gamma_{\text{DSR}}$, which is derived from the geometric clock defined by the Finsler function \eqref{f_function1}, and which obeys the principle of relativity through the deformation of the Lorentz transformation. This verification is particularly important when analyzing the effects of the Finsler deformed relativistic approach on the phenomenology of the time dilation of particle's lifetime in accelerators. In Fig.~\ref{lifetime1} we show how the lifetime of a charged pion (as a function of the energy in the lab frame), whose mass is around $140\, \text{MeV}$, would be dilated in the lab frame, when the particle propagates through the effective Finsler spacetime, if we assume that the lifetimes are connected by the clock postulate and the energy dependence is determined by the deformed Lorentz symmetry.

Note that a result that shares some similarities with this approach would be found if one had considered a different strategy involving the identification of an ``energy-momentum-dependent mass'' $m_{\text{LIV}}(p)$, that is read from the MDR, as $p_0^2-\delta^{ij}p_ip_j=m^2+\ell p_0\delta^{ij}p_ip_j\doteq m_{\text{LIV}}^2$. Taking this relation into account, it would be possible to define a modified Lorentz factor of the form $\gamma_{\text{LIV}}\doteq p_0/m_{\text{LIV}}\approx p_0/m(1-\frac{\ell}{2}\frac{p_0^3}{m^2})$. The latter kind of deformed Lorentz factor was considered in \cite{PierreAuger:2021mve} to constrain quantum gravity effects from cosmic rays, by directly inserting it as a factor that dilates the lifetime of particles at rest $\gamma_{\text{LIV}}\Delta \tau$, furnishing constraints on the quantum gravity length scale $\ell \in (-5.95\times 10^{-6},10^{-1})\ell_{\text{Planck}}$, from the MDR. 
Exactly this constraint does not translate directly to our case, since the full shape of the deformed Lorentz factor derived from the energy-dependent mass does not share exactly the same factors as the one we derive from the geometric clock \eqref{clock}, defined by the Finsler function. Further investigations would need to be carried out in order to properly compare the two proposals. Moreover, the derivation of the deformed time dilation factor from the clock postulate defining Finsler function stems from first principles, namely the time measurement. The derivation via the introduction of an ``energy-momentum-dependent mass'' $m_{\text{LIV}}(p)$ is more ad hoc and, in our opinion, as it is not immediately clear how it relates to an observer's measurement of time and the geometric clock, from which we started. Nevertheless, the two formulas share a lot of similarities, which could simplify a comparative study of the results stemming from Finsler geometry and the one from \cite{PierreAuger:2021mve}.

We can now further investigate the consequences of the modified lifetimes of particles within this framework to analyze the kinematics of particle decays in different frames, as done in Section~\ref{sec:particle_decay}. To this end, we require an expression for the deformed Lorentz transformations in terms of the momenta of the particles and their relative velocities $v$. By doing so, these processes will obey the clock postulate and preserve deformed Lorentz invariance.

\subsection{Lorentz transformations between arbitrary momenta}

For simplicity, let us assume the case of a spacetime with $1+1$ dimensions. We consider two inertial frames $S$ and $\tilde{S}$, that move with relative velocity $v$. We also assume that each observer assigns momenta $p_{\mu}$ and $\tilde{p}_{\mu}$ to a given particle. From this ansatz, we see that, to first order in $\ell$, the most general deformed transformation that connects these momenta $p_{\mu}\rightarrow \tilde{p}_{\mu}$ (and that reduces to (\ref{transf_0}) and (\ref{transf_1}) when $p_{\mu}=(m,0)$) must be of the form:
\begin{align}
&\tilde{p}_0=\gamma(p_0-vp_1)+\ell\left[A p_0 p_1+ Bp_1^2-\frac{1}{2}p_0^2(\gamma^2-1)(2\gamma^2-1)\right]\, ,\label{transf1}\\
&\tilde{p}_1=\gamma(p_1-vp_0)+\ell\left(p_0^2v\gamma^4 + Fp_0p_1 +Gp_1^2\right)\, ,\label{transf2}
\end{align}
where $A,\, B,\, F,\, G$ are general functions of $v$. It follows from dimensional considerations that the terms multiplying the perturbation parameter $\ell$ must be quadratic in momentum. To derive a deformed symmetry transformation enforcing the invariance of the MDR (\ref{MDR1}), we find the following necessary conditions for these functions:
\begin{align}
&B=-\frac{A}{v}-\frac{(1-v^2)^{3/2}-v^2-1}{2(1-v^2)^2}\, ,\label{B1}\\
&F=-\frac{A}{v}\, ,\label{F1}\\
&G=A-\frac{v[2v^2-(1-v^2)^{3/2}]}{2(1-v^2)^2}\, .\label{G1}
\end{align}
where $A$ is an arbitrary function of the relative velocity $v$ which parametrizes all possible consistent deformed Lorentz transformations. Observe that the transformations \eqref{transf1} and \eqref{transf2} preserve the Hamiltonian $H(p)=H(\tilde p)$, here given by $H(p)=p_0^2-p_1^2-\ell p_0p_1^2$, without referring to the MDR, defined by the level sets $H(p)=m^2$ or $H(p)=0$. Both cases, the massless and the massive one, are conserved by the deformed Lorentz transformations we are considering. Therefore, the results derived in this section also applies to transform the momenta of massless particles between different frames that move one relative to the other with speed $v$.

Setting $A=0$, for simplicity, we find the following set of DSR transformations
\begin{align}
[\Lambda(v;(p_0,p_1))]_{\mu}=\tilde{p}_{\mu}=\begin{cases}
&\tilde{p}_0=\gamma(p_0-vp_1)+\frac{\ell}{2}\left[ p_1^2\gamma(2\gamma^3-\gamma-1)-p_0^2(\gamma^2-1)(2\gamma^2-1)\right]\, ,\\
&\tilde{p}_1=\gamma(p_1-vp_0)+\ell v[p_0^2\gamma^4 - \frac{p_1^2}{2}\gamma(2\gamma^3-2\gamma -1)]\label{solved_finite_boost1}\, .
\end{cases}
\end{align}

In this notation, we are referring to $[\Lambda(v;(p_0,p_1))]_{\mu}$ as the $\mu$-component of the transformed momenta $(p_0,p_1)$ using boost parameter $v$. In this case, the Planck-scale correction in the energy transformation is isotropic, while that of the spatial momentum is anisotropic. Note also that this equation is dominated by $\gamma^4$ when $\gamma\gg 1$. The appearance of this $\gamma^4$ term is where the possibility of detection of effects emerges, as it works as an amplifier of Planck scale effects. On the other hand, in order to find the associated deformed boost generator, one must rely on the infinitesimal (in $v$) version of this transformation as discussed in~\cite{Amelino-Camelia:2010lsq,Jafari:2020ywd,Amelino-Camelia:2014rga}. They are given by (we also replace $v\rightarrow -v$):
\begin{align}
&\tilde{p}_0\approx p_0+vp_1\, ,\label{inf-boost0}\\
&\tilde{p}_1\approx p_1+vp_0-\ell v\left(p_0^2+\frac{p_1^2}{2}\right)\, .\label{inf-boost1}
\end{align}
This transformation on $p_{\mu}$ is equivalent to the infinitesimal version of the $\kappa$-Poincar\'e boost in the bicrossproduct basis. This can be verified by writing the above transformation as $\tilde{p}_{\mu}=p_{\mu}+v\xi_{\mu}$, where
\begin{align}
    \xi_{\mu}=\left(
\begin{array}{c}
p_1\\  p_0-\ell(p_0^2+p_1^2/2) \\
\end{array}
\right)\, ,
\end{align}
from which it is possible to identify the boost generator of $\kappa$-Poincar\'e algebra in this basis (see \cite[Eq.~(5)]{Amelino-Camelia:2014rga}):
\begin{equation}\label{boost-gen}
    {\cal N}=\xi_{\mu}x^{\mu}=x^0p_1+x^1p_0-\ell x^1\left(p_0^2+\frac{p_1^2}{2}\right)\, .
\end{equation} 
Also in \cite[Section V]{Amelino-Camelia:2014rga} it is demonstrated that this generator can be found from the Killing vectors of the related Finsler geometry, when one writes the $4$-velocities in terms of the momenta. One could wonder whether the boost found by an exponentiation of this generator could be compatible with the finite transformation \eqref{solved_finite_boost1}. The answer to this question is positive, and it requires the relation between the spacetime velocity between frames $v$ and the rapidity parameter with which the exponentiation is usually performed. We discuss this issue in detail in Appendix \ref{app:rapid}.

The effect of having $A\neq 0$ in the deformed Lorentz transformation would be that of mixing in a multiplicative way $p_0$ and $p_1$ both in energy and momentum boosts (since $F$ would no longer vanish). Notice that the terms in $B$ and $G$ in Eqs.\eqref{B1}, \eqref{G1} that do not depend on $A$ go to zero when $v\rightarrow 0$. We should also expect that $A$ goes to zero at a faster pace than $v$ in order to also have $B,G,F\rightarrow 0$ in this limit and reproduce the map from the rest frame to the lab. Here we choose $A=0$ to simplify the analysis and leave a detailed study of the case $A\neq0$ for ongoing and future studies.

%%%%%%%%%%%%%%%%%%%%%%%%%%%%%%%%%%%%%%%%%%%%%%%%%%%%%%%%%%%%%%%%%%

\section{Composition law}\label{sec:comp_law}
Another important piece for fulfilling the requirements of a relativistic deformed kinematics is the formulation of a compatible modified energy/momentum conservation law. This condition guarantees that all inertial observers will agree on the (im)possibility of interactions between elementary particles.

For dimensional reasons, the most general composition law at first order perturbation is\footnote{This law satisfies $p\oplus 0=p$ and $0\oplus p=0$.}
\begin{subequations}
\begin{align}
&(p\oplus q)_0=p_0+q_0+\ell(\alpha p_0 q_0+\beta p_1 q_1+\omega p_0 q_1+\eta p_1 q_0)\, ,\\
&(p\oplus q)_1=p_1+q_1+\ell(\delta p_1 q_0+\epsilon p_0 q_1+\lambda p_1 q_1+\mu p_0 q_0)\, ,
\end{align}\label{comp_law0}
\end{subequations}
where $(\alpha,\,  \beta,\, \omega,\, \eta,\, \delta,\, \epsilon,\, \lambda,\, \mu)$ are  dimensionless parameters, yet to be determined. Deformed relativistic compatibility is guaranteed if the action of the Lorentz transformation (\ref{solved_finite_boost1}) on composed momenta fulfills a relation of the form:
\begin{equation}\label{rel_cond1}
    \Lambda(v;((p\oplus q)_0,(p\oplus q)_1))=\Lambda(\mathfrak{v}_{v;q_0,q_1};(p_0,p_1))\oplus \Lambda(\mathfrak{v}_{v;p_0,p_1};(q_0,q_1))\, ,
\end{equation}
where, in general, it is possible that the boost parameters $\mathfrak{v}_{v;p_0,p_1}$ and $\mathfrak{v}_{v;q_0,q_1}$, appearing on the right-hand side of this relation, can dependent on the original boost parameter $v$ as well as the momenta $p$ and $q$ respectively. This is a feature that is called ``back-reaction'', and was originally proposed in \cite{Majid:2006xn} to ensure the relativistic nature of a composition law from the $\kappa$-Poincar\'e algebra in the bicrossproduct basis.Notice that when one assumes this redefinition of the boost parameter, this is also translated into a ``back-reacted'' Lorentz factor, for instance $\gamma(\mathfrak{v}_{v;p_0,p_1})=1/\sqrt{1-\mathfrak{v}_{v;p_0,p_1}^2}$, in the right-hand side of the composition law (the same for the $q$ momenta).

Here, the most general ``back-reacting'' parameters for first order deformations that we can use in the boosted composition law \eqref{rel_cond1} are
\begin{subequations}
\begin{align}
&\mathfrak{v}_{v;q_0,q_1}=v+\ell (H q_0+J q_1)\, ,\label{backq1}\\
&\mathfrak{v}_{v;p_0,p_1}=v+\ell (M p_0+R p_1)\, .\label{backp1}
\end{align}
\end{subequations}

This form of back-reaction acting on both entries of the composition law was also considered in \cite{Amelino-Camelia:2014rga}. We fully realize it in this paper and analyze its phenomenological consequences. The imposition of the relativistic condition for each component of (\ref{rel_cond1}) from Eqs.~(\ref{solved_finite_boost1}) and (\ref{comp_law0}) give the following set of conditions between the composition law parameters and those of the back-reaction:
\begin{subequations}
\begin{align}
&\alpha=0=\lambda\, ,\label{alpha1}\\
&\beta=-\frac{2+\gamma^3(J+R-2)+4\gamma^4-\gamma^2[4+J+R-v\gamma (H+M)]}{2(\gamma-1)}\, ,\\
&\delta=-\gamma\frac{\{1+\gamma^2(R-1)+2\gamma^3-\gamma(2+R-v\gamma M)\}}{2(\gamma-1)}\, ,\\
&\epsilon=\delta(R\rightarrow J\, ;\, M\rightarrow H)\, ,\\
&\omega=-\gamma\frac{\{M\gamma(\gamma-1)+v\gamma[2\gamma^2+\gamma(R-1)-1]\}}{2(\gamma-1)}\, ,\\
&\eta=\omega(R\rightarrow J\, ;\, M\rightarrow H)\, ,\\
&\mu=\frac{\{-2+\gamma^2(2+J+R)+4\gamma^3+\gamma[J+R-4+v\gamma(H+M)]\}}{2v}\, .\label{mu1}
\end{align}
\end{subequations}
Note that, so far, the only restriction we imposed was the choice $A=0$ in Eqs.~(\ref{B1}), (\ref{F1}), (\ref{G1}), which was responsible for removing mixed terms between energy and momentum in the Lorentz transformation.

%%%%%%%%%%%%%%%%%%%%%%%%%%%%%%%%%%%%%%%%%%%%%%%%%%%%%%%%%%%%%%%%%%

\subsection{Parity-invariant composition law}

As it can be seen from (\ref{MDR1}), the $3+1$-dimensional case is invariant under parity transformations ($k_0\rightarrow k_0$, $\vec k\rightarrow -\vec k$, where $k$ describes momenta $p$, $q$ and $p\oplus q$). For this reason, we implement this symmetry also in the $1+1$-dimensional case under consideration, so that the results can be translated to the general one. To do this, we require $\omega=\eta=\mu=0$ (the $\lambda$-term is null due to (\ref{alpha1})). This gives the following set of conditions on the back-reaction parameters found from Eqs.~(\ref{alpha1})--(\ref{mu1})
\begin{equation}\label{eq:BRparam}
J=-\frac{Hv\gamma}{1+\gamma}+\frac{1+\gamma-2\gamma^2}{\gamma}\, , \ \ \ \  M=-\frac{Rv\gamma}{\gamma-1}-v(1+2\gamma)\, .
\end{equation}
This restriction allows us to analyze two cases that are particularly interesting for their simplicity, namely the cases of either undeformed spatial momentum composition (see Section \ref{sssec:undefsm}) or undeformed energy composition (see Section \ref{sssec:undefe}). So far, we derived the relation between different composition laws and the necessary back-reaction parameters by considering finite boost transformations. Our approach on this issue complements the one performed in \cite{Amelino-Camelia:2014rga} where the infinitesimal version generated by \eqref{boost-gen} are assumed. We compare the results from the literature with ours in Appendix~\ref{app:comp-law}.

%%%%%%%%%%%%%%%%%%%%%%%%%%%%%%%%%%%%%%%%%%%%%%%%%%%%%%%%%%%%%%%%%%

\subsubsection{Undeformed spatial momentum conservation}\label{sssec:undefsm}

This case is important because of its possible applications in the study of two-body decays, when the analysis is done in the rest frame of the parent particle. It is realized when we impose  $\delta=0=\epsilon$ in \eqref{comp_law0} in addition to $\omega=\eta=\mu=0$.
It implies for the back-reaction parameters \eqref{eq:BRparam}:
\begin{equation}
R=J=\frac{1}{\gamma}-\gamma\, , \ \ \ \ H=M=-v\gamma\, ,
\end{equation}
which also leads to $\beta=1$. This way, the composition law reads
\begin{align}\label{und_mom_comp1}
\begin{cases}
(p\oplus q)_0=p_0+q_0+\ell p_1q_1\, ,\\
(p\oplus q)_1=p_1+q_1\, ,
\end{cases}
\end{align}
which is compatible with deformed Lorentz transformations $\Lambda(v,p)$ given by Eq.~(\ref{solved_finite_boost1}) for the back-reacting parameter
\begin{align}\label{backreaction1}
\mathfrak{v}_{v;k_0,k_1}=v+\ell\left[\left(\frac{1}{\gamma}-\gamma\right)k_1-v\gamma k_0\right]\, ,
\end{align}
where $k$ refers to momenta $p$ or $q$. In this case, the back-reaction acts equally on the first and second arguments of the composition law. One can also check this result by a straightforward calculation considering the infinitesimal transformations (\ref{inf-boost0}) and (\ref{inf-boost1}) and infinitesimal back-reacting parameter $\mathfrak{v}_{v;k_0,k_1}=v(1-\ell k_0)$.
\par
At first glance, one could be concerned about the presence of a commutative composition law, since this could indicate that we would be just working with special relativity in a weird set of momentum coordinates. But, as shown in Appendix \ref{app:comp-law}, that revisits the results of Section~VIII of Ref.\cite{Amelino-Camelia:2014rga} at the infinitesimal level, this is not the case. It is shown that relying just on the level of generators of the $\kappa$-Poincar\'e algebra, this same composition law is a genuine possibility. Complementary, another result shows the impossibility of having special relativistic results unless $\ell\rightarrow 0$.

%%%%%%%%%%%%%%%%%%%%%%%%%%%%%%%%%%%%%%%%%%%%%%%%%%%%%%%%%%%%%%%%%%

\subsubsection{Undeformed energy conservation}\label{sssec:undefe}

Another important case consists in maintaining energy conservation undeformed, since this preserves the known relations for momentum addition, defined from the coproduct structure of the bicrossproduct basis of the $\kappa$-Poincar\'e algebra. This condition is realized by assuming parity invariance and by requiring in addition $\beta=0$ in (\ref{comp_law0}), yielding
\begin{equation}
    H=\frac{v[\gamma^2(1+R)-1]}{\gamma(\gamma-1)}\, ,
\end{equation}
which implies in the following composition law
\begin{align}
\begin{cases}
(p\oplus q)_0=p_0+q_0\, ,\\
(p\oplus q)_1=p_1+q_1+\ell(\delta p_1q_0+\epsilon p_0q_1)\, ,
\end{cases}
\end{align}
where 
\begin{equation}
\delta=\frac{\gamma(\gamma^2+R\gamma-1)}{\gamma-1}\, , \ \ \ \ \epsilon=\frac{\gamma^3+R\gamma^2-1}{\gamma-1}\, .
\end{equation}
\par
This is the general case. By choosing $R=\gamma^{-1}-\gamma$, we obtain further simplifications which turn the composition law into the one from the bicrossproduct $\kappa$-Poincar\'e coproduct structure \cite{Gubitosi:2013rna},
\begin{align}\label{und_ene_comp2}
\begin{cases}
(p\oplus q)_0=p_0+q_0\, ,\\
(p\oplus q)_1=p_1+q_1-\ell p_0q_1\, .
\end{cases}
\end{align} 
This law is compatible with deformed Lorentz transformations $\Lambda(v,p)$ given by Eq.~(\ref{solved_finite_boost1}) for the back-reaction parameters
\begin{align}
&\mathfrak{v}_{v;q_0,q_1}=v-\ell\left[v\frac{(\gamma^2-1)}{\gamma}q_0+\left(\gamma-\frac{1}{\gamma^2}\right)q_1\right]\, ,\label{backreaction2-1}\\
&\mathfrak{v}_{v;p_0,p_1}=v-\ell\left[v\gamma p_0+\left(\gamma-\frac{1}{\gamma}\right)p_1\right]\,. \label{backreaction2-2}
\end{align}
One can also check this result by a straightforward calculation considering the infinitesimal transformations (\ref{inf-boost0}) and (\ref{inf-boost1}) and infinitesimal back-reacting parameters $\mathfrak{v}_{v;q_0,q_1}=v$ and $\mathfrak{v}_{v;p_0,p_1}=v(1-\ell p_0)$, which coincides with the back-reaction of the infinitesimal deformed Lorentz transformation extensively studied in the literature \cite{Majid:2006xn,Gubitosi:2013rna}.
\par
From the literature on the bicrossproduct basis of the $\kappa$-Poincar\'e algebra, the finite back-reaction acts just on the second entry of the composition law (which in our case would be the momenta $q$). In this case, we start with a very different finite form of Lorentz transformation, that coincides with the standard ones of \cite{Majid:1994cy}, only at first order in $v$. This kind of ambiguity between the Finsler and certain basis of $\kappa$-Poincar\'e was discussed in Ref.~\cite{Amelino-Camelia:2014rga} (which we revisit in appendix \ref{app:comp-law}). In particular we discuss the possibility of considering different back-reaction parameters.
 
%%%%%%%%%%%%%%%%%%%%%%%%%%%%%%%%%%%%%%%%%%%%%%%%%%%%%%%%%%%%%%%%%%

\section{Decay of a massive particle in its rest frame}\label{sec:particle_decay}

The discussion carried out in the previous sections find an useful application in the analysis of modified equations that govern cosmic-ray showers, due to the abundance of short-lived unstable particles being constantly produced.
\par
In this section we study the case of a parent particle of mass $M$ decaying into two particles with masses $m_p$ and $m_q$, with momenta $p$ and $q$, respectively. Concretely, let us consider the cases described in the previous section: {\it undeformed momentum} and {\it undeformed energy conservation}. Other cases should follow by applying similar procedures to deformed conservation laws $P_{\mu}\doteq (p\oplus q)_{\mu}$. In the following subsection, we refer to frame ``$\ast$'' as the one in which the parent particle (the one with momenta $P_{\mu}$) is at rest.

%%%%%%%%%%%%%%%%%%%%%%%%%%%%%%%%%%%%%%%%%%%%%%%%%%%%%%%%%%%%%%%%%%

\subsection{Momentum of decay products for undeformed momentum conservation}

From the rest frame condition and the composition law we find $0 = P_1^{\ast}= (p_1^{\ast}+q_1^{\ast})$, and thus $p_1^{\ast}=-q_1^{\ast}$. Then, the modified dispersion relation (\ref{MDR1}) and the composition law (\ref{und_mom_comp1}) imply 
\begin{align}\label{eq:MDRk}
    M^2 = (P_0^{\ast})^2= (p_0^{\ast} + q_0^{\ast} - \ell (p_1^{\ast})^2 )^2 = (p_0^{\ast})^2+(q_0^{\ast})^2+2(p_0^{\ast})(q_0^{\ast}) - 2\ell (p_1^{\ast})^2[p_0^{\ast}+q_0^{\ast}]\,.
\end{align}
Furthermore, $p$ and $q$ themselves satisfy the MDR (\ref{MDR1}), and thus one can replace $p_0$ and $q_0$ as functions of $p_1^*$, $q_1^*$, $m_p$ and $m_q$. After this substitution, \eqref{eq:MDRk} can be solved for $p_1^*$, which leads to a positive and a negative root 
\begin{align}\label{general-p-ast}
(p_1^{*})^{(\pm)}=-(q_1^*)^{(\pm)}=\pm \frac{\sqrt{M^4-2M^2(m_p^2+m_q^2)+(m_p^2-m_q^2)^2}}{2M}\, ,
\end{align}
\par
Notice that the composition law exactly compensates for the effect of the MDR, furnishing the same expression that would have been found in the framework of special relativity. 

An interesting possibility to be analyzed consists in the decay of a particle into a massless and a massive one. In this case, we can set $m_q=0$ (since this composition law is commutative, we would have found the same result had we set $m_p=0$). This way, the previous equation reduces to the following momentum of the produced massive and massless particles in the rest frame of the parent particle
\begin{equation}\label{ast_p_undef_mom1}
(p_1^{\ast})^{(\pm)}=-(q_1^*)^{(\pm)}=\pm\frac{M}{2}\left(1-\frac{m_p^2}{M^2}\right)\, .
\end{equation}

Although this result coincides with the one derived from special relativity, the effect of the deformed boost transformation on each decay product will furnish important corrections in the analysis to be done in the next sections.

%%%%%%%%%%%%%%%%%%%%%%%%%%%%%%%%%%%%%%%%%%%%%%%%%%%%%%%%%%%%%%%%%%

\subsection{Momentum of decay products for undeformed energy conservation}

Now, we analyze the case that coincides with the composition law that follows from the coproduct structure of the bicrossproduct basis of $\kappa$-Poincar\'e algebra, given by Eq.~(\ref{und_ene_comp2}). Again, we look at the rest frame condition from $P_1^{\ast}=0$, and now find a non-trivial relation between the momenta of the descendant particles with masses $m_p$ and $m_q$ as well as momenta $p$ and $q$. Using the on shell dispersion relation (\ref{MDR1}), we deduce to first order in $\ell$
\begin{equation}
    p_1^{\ast}=-q_1^{\ast}+\ell (q_1^{\ast})(p_0^{\ast})\approx -q_1^{\ast}+\ell (q_1^{\ast})\sqrt{m_p^2+(p_1^{\ast})^2}\approx -q_1^{\ast}+\ell (q_1)^{\ast}\sqrt{m_p^2+(q_1^{\ast})^2}\,. \label{ast2}
\end{equation}
Using again the MDR (\ref{MDR1}) for the momenta $p$, $q$, and $P_0^*$, i.e., the equation $(P_0^{\ast})^2=M^2=(p_0^{\ast})^2+(q_0^{\ast})^2+2(p_0^{\ast})(q_0^{\ast})$, we can write the energies $p_0^{\ast}$ and $q_0^{\ast}$ as a function of the spatial momenta $p_1^{\ast}$ and $q_1^{\ast}$ and the masses $m_p$ and $m_q$. In addition, with Eq.~(\ref{ast2}) we can finally express $p_1^{\ast}$ and $q_1^{\ast}$ as functions of the masses alone
\begin{subequations}
\begin{align}
q_1^{\ast}&=\pm\frac{\sqrt{M^4-2M^2(m_p^2+m_q^2)+(m_p^2-m_q^2)^2}}{2M}\, ,\label{general-p-ast2}\\
p_1^{\ast}&=\mp\frac{\sqrt{M^4-2M^2(m_p^2+m_q^2)+(m_p^2-m_q^2)^2}}{2M}\left[1-\frac{\ell (M^2+m_p^2-m_q^2)}{2M}\right]\, .\label{general-q-ast2}
\end{align}
\end{subequations}
\par
We find corrections only for the momentum $p_1$. the first one in this composition law. In fact, since the composition law is non-commutative, we can have two distinct cases depending on the order in which the particles' momenta are introduced in the deformed sum. To illustrate this issue, let us once again consider the case of the decay into a massive and a massless particle. If the massless particle is the first one in the composition law $p\oplus q$, i.e., if $m_p=0$, we find the following relations:
\begin{align}
m_p=0\Rightarrow \begin{cases}
p_1^{\ast}=\mp\frac{M}{2}\left(1-\frac{m_q^2}{M^2}\right)\left[1-\frac{\ell}{2M}(M^2-m_q^2)\right]\, ,\label{decay1}\\
q_1^{\ast}=\pm\frac{M}{2}\left(1-\frac{m_q^2}{M^2}\right)\, .
\end{cases}
\end{align}

On the other hand, if the massless one is the second particle, $m_q=0$, we find
\begin{align}
m_q=0\Rightarrow \begin{cases}
p_1^{\ast}=\mp\frac{M}{2}\left(1-\frac{m_p^2}{M^2}\right)\left[1-\frac{\ell}{2M}(M^2+m_p^2)\right]\, ,\\
q_1^{\ast}=\pm\frac{M}{2}\left(1-\frac{m_p^2}{M^2}\right)\, .\label{decay2}
\end{cases}
\end{align}
\par
So, for instance, consider a  pion decaying into a muon and a neutrino $\pi^{\pm}\rightarrow \mu^{\pm}+\nu_{\mu}(\bar{\nu}_{\mu})$, (for simplicity we shall refer these processes simply as $\pi\rightarrow \mu+\nu$). If we have the composition law of the form $p_{\pi}=p_{\mu}\oplus p_{\nu}$, where $p_{\pi}$, $p_{\mu}$ and $p_{\nu}$ refer to the energy/momentum of each of these particles, then the spatial momenta of the muon and the neutrino are given by the first and second expressions of (\ref{decay2}), respectively. This means that only the relation between the momentum of the muon, its mass, and the pion mass gets corrected in this frame. On the other hand, if we express the conservation law of this decay as $p_{\pi}=p_{\nu}\oplus p_{\mu}$, then the spatial momenta of the neutrino and the muon are given by the first and second expressions of (\ref{decay1}), respectively. This means that only the relation between the momentum of the neutrino, the mass of the muon and the mass of the pion gets corrected in this frame. Such situation does not happen in the previous subsection (undeformed momentum conservation) due to the commutativity of that composition law.

%%%%%%%%%%%%%%%%%%%%%%%%%%%%%%%%%%%%%%%%%%%%%%%%%%%%%%%%%%%%%%%%%%

\subsection{Deformed distributions for muons and neutrinos from pion decays}\label{sec:def-dist}

From the discussion in the previous subsection, we can derive some important consequences for the decay distribution of particles. The starting point is the deformed Lorentz transformation (\ref{solved_finite_boost1}) in which $\gamma$ now is considered as velocity Lorentz factor that connects the rest frame ``$\ast$'' of the parent particle and the lab.

We can calculate the limits in the laboratory frame of the energies $p_0$ and $q_0$ of secondary particles in a two-body decay. To do that, we shall connect the rest frame of the parent particle to the lab frame and realize that, since this particle's momentum is boosted by parameter $v$ with respect to the lab frame, we shall boost the produced particles momenta $p_{\mu}$ and $q_{\mu}$ with the back-reacted parameters $\mathfrak{v}_{v;p_0,p_1}$ and $\mathfrak{v}_{v;q_0,q_1}$ of Eqs.~(\ref{backp1}) and (\ref{backq1}). To determine $v$, we first rely on Eq.\eqref{transf_0}, \eqref{transf_1}, to connect the momenta $(k_0,k_1)$ of the parent particle from its rest frame $(M,0)$, also called the center of momentum (CM) frame, to the lab frame $(E_L,P_L)$:
\begin{align}
    &E_L=\gamma M-\frac{\ell}{2}M^2(\gamma^2-1)(2\gamma^2-1)\, ,\\
&P_L=-v \gamma M+\ell M^2 v_i \gamma^4\,
\end{align}
where $\gamma=1/\sqrt{1-v^2}$. By solving these equations for $v$ and using the MDR: $E_L^2-P_L^2-\ell E_L P_L^2=M^2$, we find an expression that is valid for all cases considered in this paper:
\begin{equation}
v=-\frac{P_L}{E_L}-\ell \left(P_L+\frac{P_L^3}{2E_L^2}\right)\, .\label{important-v}
\end{equation}
\par
To find the maximum and minimum energies attainable for each particle, one proceeds as follows, see appendix \ref{app:CM} for the details of the derivations:
\begin{enumerate}
    \item Calculate the maximum and minimum spatial momenta $((p_1^*)^{(\pm)},(q_1^*)^{(\pm)})$ of each particle in the CM frame. For instance, in the cases discussed in this paper, they are given by Eqs.\eqref{general-p-ast}, \eqref{general-p-ast2}, \eqref{general-q-ast2}.
    \item Calculate the energies $(p_0^*,q_0^*)$ from the corresponding MDR of each particle.
    \item Calculate the back-reaction parameters $(\mathfrak{v}_{v;p_0^*,p_1^*},\mathfrak{v}_{v;q_0^*,q_1^*})$ and the corresponding Lorentz factors.
    \item For each particle, the maximum energy is determined by boosting the energy of the particle with the largest spatial momentum. Applying the finite-boosted energy given by the first equation of \eqref{solved_finite_boost1}, we have
    \begin{align}
        p^{(\pm)}_{0}= [\Lambda(\mathfrak{v}_{v;q_0^*,(q_1^*)^{(\mp)}};(p^*_0,(p_1^*)^{(\pm)}))]_{0}\, ,\label{dE1}\\
        q^{(\pm)}_{0}= [\Lambda(\mathfrak{v}_{v;p_0^*,(p_1^*)^{(\mp)}};(q_0^*,(q_1^*)^{(\mp)}))]_{0}\, .\label{dE2}
    \end{align}
    % the boost transformation done on the energy of the particle that has the largest spatial momentum. And the minimum on the smallest spatial momentum. 
\end{enumerate}
For instance, considering the undeformed momentum conservation case, the maximum and minimum energies are found by boosting the energy of the particle using, respectively, the positive and negative roots (and its corresponding energy, found from the MDR related to each mass $m_{p,q}$) of~\eqref{general-p-ast}.
    
The expressions \eqref{dE1} and \eqref{dE2} allow us to calculate the energy differences $\Delta p_0=p_0^{(+)}-p_0^{(-)}$ and $\Delta q_0=q_0^{(+)}-q_0^{(-)}$, which allows us to find the normalized distributions of particles' energies for a product of two-body decay of an unpolarized parent
\begin{equation}
\frac{dn_p}{dp_0}=\frac{1}{\Delta p_0}\, ,\qquad \frac{dn_q}{dq_0}=\frac{1}{\Delta q_0}\, ,\label{dist1}
\end{equation}
\par
Our case study consists in the decay of a massive particle into another massive particle and a massless one, so, as a prototype that simplifies the notation, we shall refer to the {\it parent particle as a pion ($\pi$), the produced massive particle as the muon ($\mu$) and the massless one as a neutrino $\nu$}. Although neutrinos have non-vanishing masses, the sum of the masses of the three neutrino flavours are strongly constrained from CMB measurements ($\sum_{\nu}m_{\nu}<0.12\, \text{eV}$~\cite{Planck:2018vyg,Zyla:2020zbs}), which are more stringent than bounds from laboratory experiments \cite{Assamagan:1995wb}. From Eq.~\eqref{general-p-ast}, this leads us to expect that assuming a non-null neutrino mass in the analysis of this paper would produce fixed corrections of order $(m_{\nu}/m_{\pi})^2<7.3\times 10^{-19}$. Since this correction is fixed and does not present any amplifying factor, it should not interfere in the analysis carried out. For this reason, we perform our calculations assuming neutrinos as massless particles. Also, we continue referring to the parent particle momentum and energy in the lab frame as $P_L$ and $E_L$, and we refer to the energy of the produced particles as $E_{\mu,\nu}$, mass of the muon as $m_{\mu}$ and mass of the pion as $M$.

%%%%%%%%%%%%%%%%%%%%%%%%%%%%%%%%%%%%%%%%%%%%%%%%%%%%%%%%%%%%%%%%%%
\subsubsection{Undeformed momentum conservation}

As a first case, we consider the undeformed momentum conservation, that does not present ambiguities regarding the order in which the momenta are inserted in the composition law. We use Eq.~(\ref{ast_p_undef_mom1}) and the MDR (\ref{MDR1}) to find the energy of each produced particle in the frame (``$\ast$''), which gives
\begin{equation}\label{energy_star_frame}
     E_{\mu,\nu}^{\ast}=\frac{M^2\pm m_{\mu}^2}{2M}+\ell\frac{(M^2-m_{\mu}^2)^2}{8M^2}\, ,
\end{equation}
where the plus and minus signs refer to muon and neutrino energies, respectively. To calculate the maximum energy of the muon $E_{\mu}^{(+)}$ in the lab frame, we will boost the energy $E^*_{\mu}$ of \eqref{energy_star_frame}, using the maximum momentum $(p_1^{\ast})^{(+)}$ of \eqref{ast_p_undef_mom1}. To find the back-reaction boost parameter, we use $v$ of Eq.\eqref{important-v}, the momentum of the other particle $(q_1^*)^{(-)}$ of \eqref{ast_p_undef_mom1} and $E^*_{\nu}$ of \eqref{energy_star_frame} as the ingredients to determine $\mathfrak{v}_{v;E_{\nu},(q_1^*)^{(-)}}$ and its corresponding Lorentz factor. The deformed boost transformation is given by \eqref{dE1} or the first equation of \eqref{solved_finite_boost1}. The resulting equation is
\begin{align}
    E_{\mu}^{(+)}=\frac{1}{2} \left(\frac{m_{\mu }^2 \left(E_L-P_L\right)}{M^2}+E_L+P_L\right)+\ell\frac{\left(M^2-m_{\mu }^2\right)}{8M^4}\left\{m_{\mu }^2 \left[2 E_L \left(P_L-E_L\right)+M^2\right]\right.\nonumber\\
   + \left.2 M^2 E_L P_L+M^4\right\}\, .\label{und-mom-muon-max-energy}
\end{align}
\par
Similarly, the minimum energy of the muon $E_{\mu}^{(-)}$ is found from the same steps above, except that we use the minimum momentum $(p_1^{\ast})^{(-)}$ of \eqref{ast_p_undef_mom1}, instead of the maximum, and the back-reaction parameter is determined by the other particle momentum, which now reads $(q_1^{\ast})^{(+)}$. The resulting equation is
\begin{align}
  E_{\mu}^{(-)}=\frac{1}{2} \left(\frac{m_{\mu}^2 (E_L+P_L)}{M^2}+E_L-P_L\right)+\frac{\ell(M^2-m_{\mu}^2)}{8M^4}\left\{m_{\mu}^2 \left[M^2-2 E_L (E_L+P_L)\right]\right.\nonumber\\
  \left.+M^4-2 M^2 E_L P_L\right\}\, .\label{und-mom-muon-min-energy}
\end{align}
\par
For the neutrino maximum energy, we must use the momentum $(q_1^{\ast})^{(+)}$ of \eqref{ast_p_undef_mom1}, the energy $E_{\nu}^*$ of \eqref{energy_star_frame}. We also use the boost $v$ of \eqref{important-v}, the momenta of the other particle as $(p_1^*)^{(-)}$ of \eqref{ast_p_undef_mom1} and $E^*_{\mu}$ of \eqref{energy_star_frame}, in order to derive the back-reaction parameter and its Lorentz factor. Using these ingredients in \eqref{dE2} and the first equation of \eqref{solved_finite_boost1}, we find
\begin{align}
   E_{\nu}^{(+)}= \frac{(E_L+P_L) \left(M^2-m_{\mu}^2\right)}{2 M^2}+\frac{\ell (E_L+P_L) \left(M^2-m_{\mu}^2\right)}{8 M^4} &\left[2 E_L^2 P_L+E_L \left(M^2-m_{\mu}^2-2 P_L^2\right)\right.\nonumber \\
   &\left.-P_L \left(M^2+m_{\mu}^2\right)\right]\, ,\label{und-mom-neut-max-energy}
\end{align}
\par
The minimum energy of the neutrino is found by relying on $(q_1^{\ast})^{(-)}$ of \eqref{ast_p_undef_mom1} and also the energy $E_{\nu}^*$ of \eqref{energy_star_frame}. For the back-reaction, we use $v$ of \eqref{important-v}, the opposite momenta of the other particle $(p_1^*)^{(+)}$ of \eqref{ast_p_undef_mom1} and $E^*_{\mu}$ of \eqref{energy_star_frame}. This gives the following result, when applied to \eqref{dE2} and the first equation of \eqref{solved_finite_boost1}:
\begin{align}
    E_{\nu}^{(-)}=\frac{(M^2 - m_{\mu}^2)}{2M^2} (E_L - P_L)+\frac{\ell(M^2-m_{\mu}^2)}{8M^2(E_L+P_L)}&\left[E_L^3 + P_L (m_{\mu}^2 - P_L^2)- P_L (M^2 + P_L^2) \right.\nonumber\\
    &\left.- E_L (m_{\mu}^2 + P_L^2))\right]\, ,\label{und-mom-neut-min-energy}
\end{align}
\par
Most interestingly, these results found from the technique described in appendix \ref{app:CM}, can be equivalently found by calculating these thresholds in the lab frame, without relying to any frame transformation. This is the reason why there is no amplifying factors of order $E_L^3/M^2$ that are characteristic of the new deformed boost transformations (the $E_L^3$ term that appears in \eqref{und-mom-neut-min-energy} is compensated by its denominator of order $E_L$). For instance, a possible procedure to find these expressions in the lab frame consists in isolating the momenta of the neutrino $q_{\mu}=(E_{\nu},q_1)$, in terms of the momenta of the muon $p_{\mu}=(E_{\mu},p_1)$ and that of the pion $(E_L,P_L)$, like $E_{\nu}=E_L-E_{\mu}$ and $q_1=P_L-p_1-\ell p_1(P_L-p_1)$, then using the MDR, which gives
\begin{equation}\label{lab-frame-eq}
   0= E_{\nu}^2-q_1^2-\ell E_{\nu}q_1^2=m_{\mu}^2+M^2+2(p_1 P_L-E_{\mu}E_L)+\ell(E_L p_1^2+E_{\mu}P_L^2)\, .
\end{equation}

At this point, the MDR should be used to write the spatial momenta $p_1$ and $P_L$ in terms of the energies and masses of the muon, $p_0$, and of the pion, $E_L$. The solution of this equation for $p_0$ gives the two possible energies of the muon in the lab frame, which are equivalent to \eqref{und-mom-muon-max-energy} and \eqref{und-mom-muon-min-energy} for on-shell particles. And from the composition law, one can find the two possible energies of the neutrino as well, given by Eqs.\eqref{und-mom-neut-min-energy} and \eqref{und-mom-neut-max-energy}, respectively.
\par
The distributions of muons and neutrinos are found from Eqs.\eqref{dist1} as 
\begin{align}
\pi\rightarrow \nu+\mu\Rightarrow \begin{cases}
\frac{dn_{\mu}}{dE_{\mu}}&=\frac{1}{(1-m_{\mu}^2/M^2)P_L}\left[1-\ell\frac{E_L \left(M^2+m_{\mu}^2\right)}{2 M^2 }\right]\, ,\\
\frac{dn_{\nu}}{dE_{\nu}}&=\frac{1}{(1-m_{\mu}^2/M^2)P_L}-\ell\frac{E_L}{2P_L}\, .
\end{cases}
\end{align}

The flat probability distribution for detecting muons in a certain energy range from the decay of pions is modified in due to Planck-scale effects, which could be enlarged or reduced, depending on the sign of the Planck scale parameter $\ell$, which, physically, is also related to modelling superluminal ($\ell>0$) or subluminal ($\ell<0$) propagation of massless particles. To consider the decay of a neutral pion into a pair of photons, Planck-scale corrections would be found in the UV regime, as can be seen from the limit $m_{\mu}\rightarrow 0$ (besides multiplication of the distribution by 2 due to two decay products). Notice that the approach followed in this section can be generalized to the case of a decay into any two massive particles, the difference relying now in the use of Eq.~(\ref{general-p-ast}) in the MDR (\ref{MDR1}) to find the energy of each particle and their corresponding back-reaction parameters.
\par
These corrections do not present the amplification of the deformed boost transformation, i.e., of order $\ell E_L^3/M^2$ because it relies of the possible energies of particles in the lab frame. As stated above, although we used as a technique to discover these energies a transformation from the CM frame to the lab, it would also be possible to do so by calculating them directly in the lab frame (whose corrections would be due to the MDR, that does not present any amplification by itself). Since this approach is relativistic, as demonstrated in previous sections, the results found by both techniques must give the same result, and this section served also as a test of validity of the DSR nature of the undeformed momentum composition proposal.
%%%%%%%%%%%%%%%%%%%%%%%%%%%%%%%%%%%%%%%%%%%%%%%%%%%%%%%%%%%%%%%%%%
\subsubsection{Undeformed energy conservation with process $\pi\rightarrow \nu+\mu$}\label{sec:uv-und-ene1}
Now, as a first application of the undeformed energy conservation, if we set $m_p=0$, i.e, the massless particle is the first one in the non-commutative composition law of the bicrossproduct basis of $\kappa$-Poincar\'e (\ref{und_ene_comp2}), which implies that the relation between the momentum of the massive particle, its mass and the parent particle's mass is unaltered ($q_1^{\ast}$ in Eq.~(\ref{decay1})), while the momentum of the massless particle is in fact modified and given by $p_1^*$ in (\ref{decay1}).
\par
To find the threshold energies of the neutrino as the first particle in the composition law, we boost $p^*$, whose spatial component is given the first equation of \eqref{decay1}. The maximum energy of the neutrino is found by considering the positive root $p_1^*>0$ of \eqref{decay1} as the neutrino momentum $(p_1^*)^{(+)}$, calculate the energy from the MDR as $E_{\nu}^*=\left(M-m_{\mu}^2/M\right)/2-\ell \left(M^2-m_{\mu}^2\right)^2/(8 M^2)$. For the back-reaction, we use the negative root of $q_1^*<0$ in \eqref{decay1} as the muon momentum $(q_1^*)^{(-)}$, with energy in the CM frame $E_{\mu}^*= \left(M+m_{\mu}^2/M\right)/2+\ell \left(M^2-m_{\mu}^2\right)^2/(8 M^2)$. The parameter $v$ is given by \eqref{important-v}, and the back-reaction given by \eqref{backreaction2-1} and \eqref{backreaction2-2}. By performing the boost considering these ingredients, we find
\begin{align}
    E_{\nu}^{(+)}=\frac{\left(M^2-m_{\mu}^2\right) (E_L+P_L)}{2 M^2}+\ell\frac{\left(m_{\mu}^2-M^2\right)}{8 M^4}\left[m_{\mu}^2 \left(M^2-2 E_L (E_L+P_L)\right)\right.\nonumber\\
   \left. +M^4-2 M^2 E_L P_L\right]\, .
\end{align}
\par
The minimum neutrino energy is found by changing the plus to the minus signs in of \eqref{decay1} as inputs. The result is
\begin{align}
     E_{\nu}^{(-)}=\frac{\left(M^2-m_{\mu}^2\right) (E_L-P_L)}{2 M^2}+\ell\frac{\left(m_{\mu}^2-M^2\right)}{8 M^4}\left[m_{\mu}^2 \left(M^2+2 E_L (P_L-E_L)\right)\right.\nonumber\\
  \left.+M^4+2 M^2 E_L P_L\right]\, .
\end{align}
\par
The minimum energy for the muon can be found, similarly, by boosting the momentum $q_1^*<0$ of \eqref{decay1} as $(q_1^*)^{(-)}$ and its corresponding energy. And for the back-reaction, one considers the neutrino maximum momentum $(p_1^*)^{(+)}$ and its corresponding energy. For on-shell particles, the result is
\begin{align}
   E_{\mu}^{(-)}= \frac{1}{2} \left(\frac{m_{\mu}^2 (E_L+P_L)}{M^2}+E_L-P_L\right)+\frac{\ell}{8 M^4}\left[m_{\mu}^4 \left(2 E_L (E_L+P_L)-M^2\right)\right.\nonumber\\
\left.-2 m_{\mu}^2 M^2 E_L^2+M^6-2 M^4 E_L P_L\right]\, .
\end{align}
\par
The maximum muon energy is found by boosting the momentum $(q_1^*)^{(+)}$ in \eqref{decay1}, with back-reaction $(p_1^*)^{(-)}$ and considering their corresponding energies:
\begin{align}
   E_{\mu}^{(+)}= \frac{1}{2} \left(\frac{m_{\mu}^2 (E_L-P_L)}{M^2}+E_L+P_L\right)+\ell\frac{\left(M^2-m_{\mu}^2\right)}{8 M^4}\left[m_{\mu}^2 \left(M^2+2 E_L (P_L-E_L)\right)\right.\nonumber\\
  \left. +M^4+2 M^2 E_L P_L\right]\, .
\end{align}

These results are coherent with the undeformed energy conservation, since we verify that $E_{\mu}^{(-)}+E_{\nu}^{(+)}=E_L=E_{\mu}^{(+)}+E_{\nu}^{(-)}$. The distributions are found from Eq.\eqref{dist1}
\begin{equation}\label{dist_nupi}
   \pi\rightarrow \nu+\pi\Rightarrow \frac{dn_{\mu}}{dE_{\mu}}=  \frac{dn_{\nu}}{dE_{\nu}}=\frac{1}{(1-m_{\mu}^2/M^2)P_L}\left[1-\ell\frac{E_L}{2M^2}(M^2+m_{\mu}^2)\right]\, .
\end{equation}
And they read the same because of the preserved energy conservation $E_{\mu}^{(+)}+E_{\nu}^{(-)}=E_L=E_{\mu}^{(-)}+E_{\nu}^{(+)}\Rightarrow E_{\mu}^{(+)}-E_{\mu}^{(-)}=E_{\nu}^{(+)}-E_{\nu}^{(-)}$.

%%%%%%%%%%%%%%%%%%%%%%%%%%%%%%%%%%%%%%%%%%%%%%%%%%%%%%%%%%%%%%%%%%%%%%%%%%%%%%%%%%%%%%%
\subsubsection{Undeformed energy conservation with process $\pi\rightarrow \mu+\nu$}
Now, let us consider that $m_q=0$, i.e., the first particle is the muon and the second one in the composition law is the neutrino. The maximum muon energy is found by boosting the energy $E^*_{\mu}$ and the positive root of \eqref{decay2} for $p_1^*$ as $(p_1^*)^{(+)}$. As usual, for the back-reaction, we use \eqref{important-v}, the momentum given by the negative root of $q_1^*$ in \eqref{decay2} as $(q_1^*)^{(-)}$. The resulting expression is
\begin{align}
E_{\mu}^{(+)}=\frac{E_L \left(M^2+m_{\mu}^2\right)+P_L \left(M^2-m_{\mu}^2\right)}{2 M^2}-\ell\frac{\left(M^2-m_{\mu}^2\right)}{8 M^4}\left[M^2\left(m_{\mu}^2-2 E_L P_L\right)\right.\nonumber\\
\left.+2 E_L m_{\mu}^2 (P_L-E_L)+M^4\right]\, .
\end{align}

To find the minimum energy, we consider the zeroth component of the boosted momenta, starting from the center of momentum frame and considering the negative root $p_1^*$ of \eqref{decay2} as the spatial momentum (which we call $(p_1^*)^{(-)}$) and its corresponding energy (which is found from the MDR) as the components of the momenta that is being boosted. For the back-reaction, besides the same $v$ as above, we use the positive root of $q_1^*$ in \eqref{decay2} as $(q_1^*)^{(+)}$ and its energy. The result is
\begin{align}
  E_{\mu}^{(-)} =\frac{E_L \left(M^2+m_{\mu}^2\right)+P_L \left(m_{\mu}^2-M^2\right)}{2 M^2}-\ell\frac{\left(M^2-m_{\mu}^2\right) }{8 M^4}\left[M^2 \left(2 E_L P_L+m_{\mu}^2\right)\right.\nonumber\\
 \left.-2 E_L m_{\mu}^2 (E_L+P_L)+M^4\right]\, .
\end{align}

For the maximum neutrino energy, we boost $E_{\nu}^*$ with momentum $(q_1^*)^{(+)}$ from the positive root of \eqref{decay2}. For the back-reaction, besides $v$ of \eqref{important-v}, we use energy $E_{\mu}^*$ and momentum $(p_1^*)^{(-)}$ as the negative root in \eqref{decay2}. We find
\begin{equation}
    E_{\nu}^{(+)}=\frac{\left(M^2-m_{\mu}^2\right) (E_L+P_L)}{2 M^2}+\ell\frac{\left(M^2-m_{\mu}^2\right) \left[(M^2-m_{\mu}^2)(E_L+P_L)^2-2P_L^2M^2\right]}{2 M^4}\, .
\end{equation}

The minimum neutrino energy is found by boosting the energy $E_{\nu}^*$, the momentum $(q_1^*)^{(-)}$ from the negative root of $q_1^*$ in \eqref{decay2}. For the back-reaction on the boost parameter, we consider the same $v$ above, $E_{\mu}^*$ and the opposite momentum $(p_1^*)^{(+)}$ from the positive root of $p_1^*$ in \eqref{decay2}:
\begin{align}
    E_{\nu}^{(-)}=\frac{(E_L-P_L) \left(M^2-m_{\mu}^2\right)}{2 M^2}+\ell\frac{\left(M^2-m_{\mu}^2\right)}{8 M^2 (E_L+P_L)}\left[E_L \left(M^2-m_{\mu}^2-2 P_L^2\right)\right.\nonumber\\
 \left.+P_L \left(-M^2+m_{\mu}^2-2 P_L^2\right)\right]\, .
\end{align}

The distributions are also the same for both particles due to the undeformed conservation law and they read
\begin{equation}
    \pi\rightarrow \mu+\nu\Rightarrow \frac{dn_{\mu}}{dE_{\mu}}=\frac{dn_{\nu}}{dE_{\nu}}=\frac{1}{(1-m_{\mu}^2/M^2)P_L}-\ell \frac{E_L}{2P_L}\, .
\end{equation}
Here, we notice the difference in comparison with the previous result \eqref{dist_nupi} due to the non-commutativity of the composition law.

%%%%%%%%%%%%%%%%%%%%%%%%%%%%%%%%%%%%%%%%%%%%%%%%%%%%%%%%%%%%%%%%%%%%%%%%%%%%%%%%%%%%%%

\subsection{Some phenomenological consequences}
From the analysis presented, we can draw some conclusions that might guide future investigations in this field when considering, for instance, particles produced in cosmic-ray showers in the atmosphere. In this case, there are three main points that gain corrections and that are important for calculating the spectra of the decay products \cite{gaisser2016a}:

\medskip
\noindent\textbf{1. Deformed lifetimes of the particles, that furnishes the distance they travel.}\\
The deformed lifetimes were discussed in this Finsler approach in Ref.~\cite{Lobo:2020qoa} and was also recapped in this paper in Eq.~(\ref{eq:lifetime}).

\medskip
\noindent\textbf{2. Deformed distributions of particles.}\\
The deformed distributions of particles were analyzed for the case of a two-body decay (in which one of the particles is massive) in the previous sections. From the discussion of Section \ref{sec:def-dist}, we verified the presence of Planck-scale  dimensionless corrections of order $\ell E_L$ which would be unobservable is $\ell$ is of the order of the Planck length. This result is compatible with the relativistic nature of our approach, since the same expressions are found by performing calculations in the lab frame directly, and in this frame, the only mathematical inputs are the modified composition law and the MDR. For this reason, this kind of correction is suppressed compared to those of order $\ell E_L^3/M^2$ that appear in the deformed lifetime due to the action of deformed Lorentz transformation and the deformed clock considered previously. We stress that this is not a property of the specific kinds of composition laws and MDRs considered, but are instead a manifestation of the relativistic nature of this approach.

\medskip
\noindent\textbf{3. Deformed minimum and maximum attainable energies of each produced particle and, consequently, of the parent particle, for integration purposes.}\\
The maximum and minimum energies for each particle are determined by the ``$(+)$'' and ``$(-)$'' signs in Eqs.~(\ref{dE1}) and (\ref{dE2}), as was calculated in the previous section. Our analysis was done considering $1+1$D, which means that we have a discrete set of possible energies $E_{\mu,\nu}^{(\pm)}$ attainable for each particle. When one considers special relativity in $3+1$D, this discrete set is actually given by two continuous intervals that have $E_{\mu,\nu}^{(\pm)}$ as maxima and minima (when $\ell=0$) and the intermediary attainable energies are usually controlled by the cosine of the angle between some particles' spatial momenta (usually the parent and one of the produced particles, which can be seen from the derivation of Eq.\eqref{lab-frame-eq}, where the product $p_1 P_L$ would also carry a cosine function.). If one assumes that something similar also happens in the DSR case in $3+1$D, then it is expected that the intervals of possible attainable energies of the produced particles in the lab frame are given by
\begin{equation}
 E_{\mu,\nu}^{(-)}\leq E_{\mu,\nu}\leq  E_{\mu,\nu}^{(+)}\, .
 \end{equation}
 
This is important because, when integrations between intervals defined by these inequalities are to be considered, then only corrections of order $\ell E$ shall appear, which are strongly suppressed by those of order $\ell E^3/M^2$, that emerge from the dilated lifetime of particles and the longer path along which they propagate (nevertheless, if $\ell$ is of the order of the Planck length, then the term $\ell E$ would be unobservable by itself).
\par
We considered the case of the pion decay for simplicity reasons, but similar procedures could be carried out if one aims to analyze different channels, while still preserving the relativistic principle at the Planck scale. We can see, immediately that in the decay $\pi^0\rightarrow \gamma\gamma$ (which could be found by just setting $m_{\mu}\rightarrow 0$, $E_L$ as the neutral pion energy in the lab frame and multiplying the distribution by 2, since there are two identical decay products), the distributions read the same for the examples analyzed in this paper and present a minor dimensionless correction equals to $-\ell E_L/2$.

As has been highlighted in this paper, this approach is based on a preservation of the relativity principle, by a deformation of the Lorentz symmetry and the corresponding rules to compose momenta. A scenario that breaks Lorentz symmetry can be realized by relaxing at least one of the following ingredients: 
\begin{enumerate}
    \item Modified Dispersion Relation,
    \item Modified Lorentz Transformation,
    \item Modified Composition Law.
\end{enumerate}
 
Since item 1) is an input of our analysis, we shall comment on the other two points. Let us suppose that one preserves item 2) and breaks item 3). Then, if one assumes that this modified Lorentz transformation determines the dilated lifetime of particles, then a correction of the kind $\ell E_L^3/M^2$ is expected to appear. For the energy thresholds and distributions, if one performs the relevant calculations in the lab frame, the only kind of correction that would emerge would come from the MDR and usual threshold effects considered in a LIV scenario \cite{HAWC:2019gui} would emerge. On the other hand, if these calculations is done in the center of momentum frame and the results are transformed to the lab frame, then an extra amplification of order $\ell E_L^3/M^2$ are expected to emerge in the energy equations.
\par
If item 2) is relaxed, and one considers an MDR with the usual Lorentz transformations to map observer frames onto each other, one immediately runs into the problem that the MDR is not invariant under these transformations. Hence one needs to fix a preferred frame in which the MDR has the desired form and derive all observables starting from this frame. Alternatively one might assume that the MDR is valid in any frame, but then, mapping from one frame to the another with the usual Lorentz transformations would yield that an observer originally in a certain frame sees a different dispersion relation, compared to one who is mapping itself into the frame with a Lorentz transformation.  This leads to different predictions for time dilations between frames depending, if they are derived from the MDR or from Lorentz transforming between the frames. If, in addition on relaxes also item 3) and consider an MDR together with usual Lorentz transformations and the usual composition law for momenta the only source of corrections would be the MDR, and threshold effects are frame dependent. One obtains different results if one derives them in a certain frame directly or if one maps them from one frame to another. If one keeps item 3) (adapted to the MDR), then we expect that threshold corrections that might appear in the lab frame to be suppressed, as is usually the case when one performs calculations based on modifying the dispersion relation and the composition law simultaneously (see, for instance Section 3 of \cite{AmelinoCamelia:2008qg}).

Another phenomenological consequence of the dilated lifetime of decaying particles in DSR are the different fluxes of particles resulting from the decay of, for instance, pions. The lifetimes of $\pi^0$ and $\pi^\pm$ would be altered according to Eq.~(\ref{eq:lifetime}), and so would the relative fluxes of neutrinos, photons, and electrons. In cosmic-ray showers, this would directly translate into a difference in the hadronic component of the shower with respect to the electromagnetic one. Naturally, other unstable particles commonly found in atmospheric showers such as kaons and eta mesons would be affected in a similar fashion.

Interestingly, the MDR presented in Eq.~(\ref{MDR1}) also has important implications for interpreting signatures of cosmic messengers related to two-body decays. For instance, the interpretation of electromagnetic observations across the whole spectrum together with a neutrino signal for the neutrino-emitting objects~\cite{IceCube:2018dnn, IceCube:2018cha} could have to be revisited, considering a possible enhancement or suppression of the neutrino flux with respect to the gamma-ray one, depending on whether we are dealing with a superluminal scenario ($\ell > 0$) or subluminal ($\ell < 0$). At ultra-high energies, this would also affect the expected fluxes of the long-sought cosmogenic neutrinos and photons, stemming from UHECR interactions with pervasive photon fields such as the CMB~\cite{AlvesBatista:2018zui}. Furthermore, depending on how much the lifetime of a particle changes compared to the special-relativistic kinematics, effects such as synchrotron emission by charged particles could become a relevant energy-loss mechanism, depending on the magnetic field.

%%%%%%%%%%%%%%%%%%%%%%%%%%%%%%%%%%%%%%%%%%%%%%%%%%%%%%%%%%%
\section{Final remarks}\label{sec:conc}
In a previous paper \cite{Lobo:2020qoa}, a finite deformed Lorentz symmetry connecting the rest and laboratory frames emerged from Finsler geometry, containing an amplifying factor, of order $\ell p_0^3/M^2$ in the dilated lifetime of the particles in the lab frame, which could lead to observations close to Planck scale sensitivity. 
%\sout{in particular in the time dilation of the lifetime of particles, that can lead to observations close to Planck scale sensitivity} 
%\textcolor{blue}{which indicates corrections in the dilated lifetime of particles in the lab frame, which could lead to observations close to Planck scale sensitivity}.

In this paper, we continued this analysis on two stages:

First, in order to have a more complete DSR framework, we generalized the $\kappa$-Poincar\'e-inspired  results of \cite{Lobo:2020qoa} by constructing finite deformed Lorentz transformations which connect the momenta of particles in two different frames (in $1+1$D) that move relatively to each other. This construction is necessary for the analysis of phenomenological possibilities of Finsler-DSR in a rich variety of contexts, like the analysis of cosmic rays.

In order to achieve the complete relativistic realization, we also built the general momentum composition law that is compatible with this finite transformation at first order in the deformation scale. To achieve this, we needed to introduce a back-reaction acting on both ``entries'' of the composition law, which means that the each boost parameter acting on momenta of a given particle depends on the momenta of the other particle. This condition is fundamental to guarantee that all inertial observers agree about the nature of the vertices of interactions between fundamental particles. In particular, we focused on the cases of undeformed spatial momentum conservation, and undeformed energy conservation that is compatible with the coproduct structure of the $\kappa$-Poincar\'e algebra in the bicrossproduct basis.

On the second stage, we applied these constructions (modified dispersion relation, Finsler-compatible deformed Lorentz transformation and composition law with back-reaction) to consider the decay of a massive parent particle into two descendant ones. We derived kinematical equations that allowed us to deduce the correction to the distributions of produced particles when one the products is massless.

We verified that these quantities carry dimensionless corrections of order $\ell E_L$ (where $\ell$ is expected to be of the order of the inverse of Planck energy and $E_L$ is the energy of the parent particle as measured in the lab frame), which would be unobservably small. The results read the same whether the energy is calculated in the lab frame or calculated in the center of momentum frame and then transformed to the lab frame. This result consists in a manifestation of the relativistic nature of this approach, in which possible amplifications from transforming to the center of momentum frame are cancelled, due the presence of a modified composition law and its back-reaction. 
%\IL{Nevertheless}, in the DSR scenario, we still have a modification due to an important amplifier of the order $\ell E_L^3/M^2$ \textcolor{blue}{in the dilated lifetime of the particles in the lab frame}. 
%Energy thresholds are also modified, but with corrections that are suppressed compared to those of the lifetime.
We expect that this should also be case even if higher orders of Planck-scale corrections are considered. In a LIV scenario, usual threshold effects that had already appeared in other approaches are expected if calculations are done in the lab frame \cite{HAWC:2019gui}. Although, if one still considers that the Lorentz transformation is deformed and  the composition law is the usual one of SR, then some amplified corrections may appear by performing calculations in different frames before transforming to the lab frame.

This articles aims at proposing a novel procedure for constraining the Planck scale, based on a connection with the particle and astroparticle physics communities, in a way that respects relativistic principles. The steps followed in this paper can be generalized for the sake of allowing the exploration of higher orders of perturbation in the Planck scale from the analysis of very-high energy particles, which complement and extend the studies started in~\cite{Carmona:2016obd}.

Moreover, it opens the opportunity for future studies of phenomenology of the DSR framework. For example combining the results on the energy dependence of dilated particle lifetimes and the results of modifications in the particle decays one may address the muon puzzle alluded to in Section~\ref{sec:intro}. A complete reinterpretation of the development of atmospheric showers within the DSR framework would allow for robust theoretical predictions that could be confronted with  data from facilities such as the Pierre Auger Observatory~\cite{PierreAuger:2015eyc} and its forthcoming enhancement dedicated to muon measurements, AMIGA~\cite{PierreAuger:2020hrz}, as well as GRANDProto300~\cite{Decoene:2019sgx}, the prototype of the Giant Radio Array for Neutrino Detection (GRAND)~\cite{GRAND:2018iaj}. Furthermore, astrophysical strategies combining multiple messengers could also be of use to test this idea using the fluxes of secondary particles stemming from the decay of muons or pions, for example.

\acknowledgments
C.P. was funded by the Deutsche Forschungsgemeinschaft (DFG, German Research Foundation) - Project Number 420243324 and acknowledges support from the excellence cluster EXC-2123 QuantumFrontiers - 390837967 funded by the Deutsche Forschungsgemeinschaft (DFG, German Research Foundation) under Germany's Excellence Strategy. I. P. L. was partially supported by the National Council for Scientific and Technological Development - CNPq grant 306414/2020-1. R. A. B. is funded by the ``la Caixa'' Foundation (ID 100010434) and the European Union's Horizon~2020 research and innovation program under the Marie Marie Sklodowska-Curie-Curie grant agreement No 847648, fellowship code LCF/BQ/PI21/11830030. V. B. B. was partially supported by the National Council for Scientific and Technological Development - CNPq grant 307211/2020-7. P. H. M. thanks Coordena\c c\~ao de Aperfei\c coamento de Pessoal de N\'ivel Superior - Brazil (CAPES) - Finance Code 001 for financial support.
The authors would like to acknowledge networking support by the COST Action QGMM (CA18108), supported by COST (European Cooperation in Science and Technology). The authors would like to thank the anonymous referees of the article, whose comments and questions helped us to improve the results.

%%%%%%%%%%%%%%%%%%%%%%%%%%%%%%%%%%%%%%%%%%%%%%%%%%%%%%%%%%%
\appendix 

\section{Relation between rapidity parameter and physical velocity between frames}\label{app:rapid}
In \eqref{boost-gen} we displayed the boost generators of $\kappa$-Poincar\'e modified Lorentz transformations. Here we will derive the finite transformations from exponentiation. We find the following result:
    \begin{align}\label{eq:finiteLT}
    \begin{split}
    \tilde{p}_0&=p_0 \cosh (\xi )+p_1 \sinh (\xi )-\ell \sinh ^2\left(\tfrac{\xi }{2}\right) \left[\left(p_0^2+p_1^2\right) \cosh (\xi )+p_0 (p_0+2 p_1 \sinh (\xi ))\right]\, ,\\
    \tilde{p}_1&=p_0 \sinh (\xi )+p_1 \cosh (\xi )-\ell \sinh \left(\tfrac{\xi }{2}\right) \left[p_0^2 \cosh \left(\tfrac{\xi }{2}\right)+\left(p_0^2+p_1^2\right) \cosh \left(\tfrac{3 \xi }{2}\right)+2 p_0 p_1 \sinh \left(\tfrac{3 \xi }{2}\right)\right]\, ,
    \end{split}
    \end{align}
    that can be found, for instance, from the first order perturbation of the transformation (28) of \cite{Gubitosi:2013rna}. To derive this result, one sets conditions on $\tilde{p_{\mu}}$ as $\tilde{p_{\mu}}(\xi=0)=p_{\mu}$ and on  $dp_{\mu}/d\xi|_{\xi=0}$ (where $\mu=0\, ,1$). However, this mathematical result does not say what is the physical meaning of $\xi$. In special relativity, we know that for frames moving relative one to the other with velocity $v$ (let us assume $v\geq 0$, for simplicity), one has $\xi=\arcsinh(v/\sqrt{1-v^2})$ (one gets a $\pm$ signs depend whether the frames are moving towards or outwards one another). In the DSR case, in principle, one cannot assume that a priori, i.e., there should have a way to link the rapidity $\xi$ with the velocity $v$. Since we are dealing with deformations of quantities governed by dimensionless contributions given in terms of the Planck length multiplied by the momenta, $\ell p_{\mu}$, also the connection between $\xi$ and $v$ may be given by a momentum-dependent Planck-scale correction. In first order in $\ell$ (considering the dimensions of the quantities involved) this should be of the form
    \begin{equation}
    \xi=\arcsinh(v/\sqrt{1-v^2}) + \ell(a p_0+b p_1)\, ,
    \end{equation}
    where $a$ and $b$ are dimensionless functions of $v$. 
    
    We can determine $a,b$ by plugging $\xi$ into \eqref{eq:finiteLT} and demanding that this finite transformations matches the transformations presented in our paper in \eqref{solved_finite_boost1} 
    \begin{eqnarray}
    \tilde{p}_0=\gamma(p_0+vp_1)+\frac{\ell}{2}\left[ p_1^2\gamma(2\gamma^3-\gamma-1)-p_0^2(\gamma^2-1)(2\gamma^2-1)\right]\, ,\\
    \tilde{p}_1=\gamma(p_1+vp_0)-\ell v[p_0^2\gamma^4 - \frac{p_1^2}{2}\gamma(2\gamma^3-2\gamma -1)]\,.
    \end{eqnarray}
    This is indeed possible and fixes $\xi$ to be
    \begin{equation}
        \xi(v)=\arcsinh(v\gamma)+\ell[-v^3\gamma^3 p_0+p_1(\gamma^3-1)]\, ,
    \end{equation}
    or inversely
    \begin{equation}
        v(\xi)=\tanh{(\xi)}+\ell \cosh{(\xi)}[p_0\tanh^3{(\xi)}+p_1(\sech^3{(\xi)-1})]\, .
    \end{equation}
    This map reduces to the identity, when we consider only first order terms in $v$ and we also have $\xi(v=0)=0$.

    So, the Finsler geometry of the spacetime probed by these particles are allowing us to express the rapidity parameter found when one exponentiates the $\kappa$-Poincar\'e boost transformation in terms of the physical spatial velocity between frames. The amplification that we found in this approach is hidden in the relation between the rapidity parameter and the physical spatial velocity between frames, that was found by relying on the Finsler nature of the spacetime probed by particles in which this MDR becomes important.
    
    %%%%%%%%%%%%%%%%%%%%%%%%%%%%%%%%%%%%%%%%%%%%%%%%%%%%%%%%%%%

\section{Revisiting some possible composition laws at the infinitesimal level in the bicrossproduct basis of $\kappa$-Poincar\'e}\label{app:comp-law}  

Our work builds up on previous approaches to the $\kappa$-Poincar\'e-inspired Finsler geometry of reference \cite{Amelino-Camelia:2014rga} by other authors. We notice that Section~VIII of that approach also calculated the modified composition law that is DSR-compatible with a generator of boosts (that includes the bicrossproduct basis of $\kappa$-Poincar\'e as a special case), along with the corresponding back-reaction parameters. We stress that this calculation was done just at the infinitesimal level in the boost parameter, which our paper extends, considering its finite version in the boost parameter (and by furnishing its connection with the velocity between frames).
    \par
    Relying on the definitions of that paper, the bicrossproduct basis of $\kappa$-Poincar\'e generator is the one that, in Eq.(114) of \cite{Amelino-Camelia:2014rga}, satisfies $\alpha=\gamma=0$ in:
    \begin{equation}\label{generator1}
        {\cal N}_{\kappa-compatible}=p_1x^0+p_0x^1+\ell(\alpha p_0p_1x^0+\gamma(p_0p_1x^1+p_1^2x^0)+(\alpha-1)p_0^2x^1-\frac{1}{2}p_1^2x^1)\, ,
    \end{equation}
    and the parity invariant modified composition law reads (Eqs.(115) and (116)):
    \begin{align}
       & (p\oplus q)_0=p_0+q_0+\ell({\cal A}p_0q_0+{\cal B}p_1q_1)\, ,\label{comp1}\\
        &(p\oplus q)_1=p_1+q_1+\ell({\cal C}p_1q_0+{\cal D}p_0q_1)\, .\label{comp2}
    \end{align}
    
    By considering the back-reaction acting on both entries of the composition law in order to consider a generalized covariance condition, one should have the composition law as given by Eq.(126)
    \begin{align}
        (p\oplus q)'_{\mu}=
        \begin{cases}
            =k'_{\mu}=k_{\mu}+\xi\{{\cal N}_{\kappa-compatible},k_{\mu}\}\, \\
            =(p'\oplus q')_{\mu}=[(p+\xi_1\{{\cal N}_{\kappa-compatible},p\})\oplus (q+\xi_2\{{\cal N}_{\kappa-compatible},q\})]_{\mu}\, ,
        \end{cases}
    \end{align}
    where the back-reaction parameters $\xi_i$ that acts on entries $1$ and $2$ are given just below these equations as $\xi_1=\xi(1+\ell(f_{11}q_0+f_{12}q_1))$ and $\xi_2=\xi(1+\ell(f_{21}p_0+f_{22}p_1))$, and $\{,\}$ are the Poisson brackets and a canonical structure for the phase space variables is assumed. This compatibility condition for DSR-covariance of this approach reads
    \begin{align}
        f_{12}&=f_{22}=\gamma\, ,\\
        {\cal A}&=0\, \\
        {\cal B}&=2\alpha-1-f{11}-f_{21}\, ,\\
        {\cal C}&=\alpha-1-f_{21}\, ,\\
        {\cal D}&=\alpha-1-f_{11}\, .
    \end{align}
    
    As can be seen, even when one assumes the generator of $\kappa$-Poincar\'e boosts in the bicrossproduct basis, which means taking $\alpha=\gamma=0$ in Eq.\eqref{generator1}, one finds an {\it ambiguity} in the definition of the deformed composition law/back-reaction. In fact, this condition would lead to the following result (using our \eqref{comp1}, \eqref{comp2})
    \begin{align}
      (p\oplus q)_0&=p_0+q_0-\ell(1+f_{11}+f_{21})p_1q_1\, ,\label{comp2-1}\\
    (p\oplus q)_1&=p_1+q_1-\ell[(1+f_{21})p_1q_0+(1+f_{11})p_0q_1)] \, ,\label{comp2-2}\\
    \xi_1&=\xi(1+\ell f_{11}q_0)\, ,\label{back1}\\
    \xi_2&=\xi(1+\ell f_{21}p_0)\, .\label{back2}
    \end{align}
    
    This way, one can see that a commutative composition law can be achieved whenever $f_{11}=f_{21}$ and one still has a genuine deformed composition law with possible back-reaction parameters. For instance, the case $f_{11}=f_{21}=-1$ gives the case of the undeformed momentum composition with its back-reaction considered in our paper. As can be seen, the commutative composition law is compatible with the fact that the back-reaction acts equally on both entries $\xi_1=\xi_2$.
    \par
    One could still have absence of back-reaction by considering $f_{11}=f_{22}=0$, at the cost of having both deformed energy and momentum conservation. The case that stems from the co-product structure in $\kappa$-Poincar\'e is given by $f_{11}=0$, $f_{21}=-1$. In any case, one can {\bf never} have the trivial case of undeformed composition rule $p\oplus q=p+q$.
    \par
    Our paper analyzed the DSR-covariance from the viewpoint of the finite boost parameters and giving the relation between the rapidity parameter and the physical velocity between frames (which is the origin of the new amplifying factor discovered).
    \par
    This representation of the boost generator of $\kappa$-Poincar\'e (given by the case $\alpha=\gamma=0$ of \eqref{generator1}) is found from the trivial (Darboux) sympletic structure of the phase space with the Casimir operator ${\cal C}=p_0^2-p_1^2-\ell p_0p_1^2$ in the same coordinate system. So from the point of view of the momentum variables, the set of coordinates are already established in these calculations. These composition laws only emerge due to ambiguities in the compatibility between the boost, composition law and back-reaction. These three items allow the existence of such an ambiguity. For this reason, even when considering the different possibilities in \eqref{comp2-1}, \eqref{comp2-2}, \eqref{back1}, \eqref{back2}, we are still under a legitimate $\kappa$-Poincar\'e approach. And the correct choice should be determined by experiments.    
    
    %%%%%%%%%%%%%%%%%%%%%%%%%%%%%%
    %%%%%%%%%%%%%%%%%%%%%%%%%%%%%

\section{Particles' momenta from the center of momentum frame to the lab frame}\label{app:CM}

In this appendix, we explain in further details the definition of the energy range and the reason why the distribution of particles can be different in this two body decay. We start by setting the following detailed notation for the Lorentz transformation and composition law with back-reaction in all generality.
\par
In this appendix, we refer to
\begin{equation}
\mathfrak{v}_{v;p_0,p_1} \qquad \text{(Back-reaction parameter)}
\end{equation}
as the back-reaction parameter that depends on the boost parameter $v$, the energy $p_0$ and the spatial momentum $p_1$. So the first entry of $\mathfrak{v}_{v;p_0,p_1}$ presents the boost parameter to which the back-reaction refers to and the other entries, after the semicolon ``;'', refer to the energy and spatial momentum of the particle that deforms $v$.
\par
We refer to
\begin{equation}
    [\Lambda(\mathfrak{v}_{v,p_0,p_1};(k_0,k_1))]_{\mu} \qquad \qquad \text{(Components of the boosted momenta)}
\end{equation}
as the $\mu$-component ($\mu=0,\, 1$) of the action of the boost $\Lambda$ on momenta $k=(k_0,k_1)$ with parameter $\mathfrak{v}_{v;p_0,p_1}$.
\par
We refer to 
\begin{align}
k_{\mu}&=[p\oplus q]_{\mu} \qquad &\text{(Composition law)}\\
    \left[\Lambda(\mathfrak{v}_{v;0,0};(k_0,k_1))\right]_{\mu}&=\left[\Lambda(\mathfrak{v}_{v;q_0,q_1};(p_0,p_1))\oplus \Lambda(\mathfrak{v}_{v;p_0,p_1};(q_0,q_1))\right]_{\mu} \qquad &\text{(Covariance condition)}
\end{align}
as the deformed composition law and DSR-covariance condition that assures that different observers agree on the conservation laws that govern interactions. Notice that, on momenta $k$, we used an undeformed boost parameter $\mathfrak{v}_{v;0,0}=v$. 
\par
At the center of momentum (CM) frame of this interaction, we have $k_1^*=(p\oplus q)_1=0$ (for which we use the ``$*$'' symbol to describe its related quantities). If we transform the composition law of a two-body decay from the center of momentum frame to the lab frame (which are connected by the boost parameter $v$) we have
\begin{equation}\label{conserv1}
    \left[\Lambda(\mathfrak{v}_{v;0,0};(m_k,0))\right]_{\mu}=\left[\Lambda(\mathfrak{v}_{v;q_0^*,q_1^*};(p^*_0,p^*_1))\oplus \Lambda(\mathfrak{v}_{v;p_0^*,p_1^*};(q_0^*,q^*_1))\right]_{\mu}\, ,
\end{equation}
where $p^*=(p_0^*,p_1^*)$, $q^*=(q_0^*,q_1^*)$ and $k^*=(m_k,0)$, where we refer to $m_k$ as the mass of the parent particle with momenta $k$ and corresponds to its energy in the rest frame.
\par
Usually, one has two real roots when solving the equations for the momenta of particles in the center of momentum condition $(p\oplus q)_1=0$. For instance, if one considers the larger root of $p_1^*=(p_1^*)^{(+)}$ and the corresponding energy $p_0^*\left((p_1^*)^{(+)}\right)$ from the MDR, then $q_1^*$ follows straightforwardly from the condition, $[(p_1^*)^{(+)}\oplus q^*]_1=0$ and we denote it by $(q_1^*)^{(-)}$, and the energy $q_0^*$ from the MDR for momenta $q$. This way, the momenta of the particles in the lab frame read
\begin{align}
p^{(+)}_{\mu}\doteq [\Lambda(\mathfrak{v}_{v;q_0^*,(q_1^*)^{(-)}};(p^*_0,(p_1^*)^{(+)}))]_{\mu}\, ,\label{p+}\\
q^{(-)}_{\mu}\doteq [\Lambda(\mathfrak{v}_{v;p_0^*,(p_1^*)^{(+)}};(q_0^*,(q_1^*)^{(-)}))]_{\mu}\, ,\label{q-}\end{align}
and this is one possible outcome when transforming from the CM frame to the lab.
\par
The other outcome follows if we consider the other set of roots with $p_1^*=(p_1^*)^{(-)}$ as the smaller root of the first particle and $q_1^*=(q_1^*)^{(+)}$ as the larger root of the second particle, from which we can find the quantities $p_0^*$, $q_1^*$, $q_0^*$ from the same procedure as above. This way, the other possibility of momenta in the lab frame reads
\begin{align}
p^{(-)}_{\mu}\doteq [\Lambda(\mathfrak{v}_{v;q_0^*,(q_1^*)^{(+)}};(p^*_0,(p_1^*)^{(-)}))]_{\mu}\, ,\label{p-}\\
q^{(+)}_{\mu}\doteq [\Lambda(\mathfrak{v}_{v;p_0^*,(p_1^*)^{(-)}};(q_0^*,(q_1^*)^{(+)}))]_{\mu}\, .\label{q+}
\end{align}
\par
In summary, the conservation law \eqref{conserv1} reads the following in these two distinct cases, in the lab frame
\begin{align}
   \left[\Lambda(\mathfrak{v}_{v;0,0};(m_k,0))\right]_{\mu}=\left[p^{(+)}\oplus q^{(-)}\right]_{\mu}\, ,\\
   \left[\Lambda(\mathfrak{v}_{v;0,0};(m_k,0))\right]_{\mu}=\left[p^{(-)}\oplus q^{(+)}\right]_{\mu}\, .
\end{align}
These are the only possibilities in $1+1$D, since the each particle can only be moving in one direction or the opposite. These $0$-components of \eqref{p+}-\eqref{q+} are the possible combination of particles' energies in the lab frame. This must give the same result whether we calculate these energies following this technique of transforming from the CM frame to the lab or calculating directly in the lab frame, since this is a relativistic framework, in fact, the verification of this equivalence is a test of validity of the DSR nature of the approach. We follow these lines, because sometimes it could be easier from a technical point of view to perform the calculation in one frame or the other. 
\par
This also allows us to calculate the energy difference 
\begin{equation}
    \Delta p_0=p^{(+)}_0-p^{(-)}_0,\qquad  \Delta q_0=q^{(+)}_0-q^{(-)}_0\, .
\end{equation}

The $(+)$ and $(-)$ energies define the range  attainable for each produced particle as measured in the lab frame in the DSR scenario.
%%%%%%%%%%%%%%%%%%%%%%%%%%%%%%%%%%%%%%%%%%%%%%%%%%%%%%%%%%%

\bibliographystyle{JHEP}
\bibliography{FiniteFinsler}

\end{document}